\documentclass[12 pt]{article}
\usepackage{latexsym}
\usepackage[tbtags]{amsmath}
\usepackage{amsmath,amssymb,graphicx}
\usepackage{color}
\usepackage{graphicx}
\usepackage{graphics}
\usepackage{authblk}
\usepackage{cite}
\usepackage[toc,page]{appendix}
\usepackage[usenames,dvipsnames]{xcolor}
\usepackage[pdfborder={0 0 0}]{hyperref}
\usepackage{algorithm} 
\usepackage{algorithmic}
\usepackage{lipsum}
\usepackage{tablefootnote}
\usepackage{fixltx2e}
\usepackage{enumitem}
\usepackage{multirow}
\usepackage{multicol}
\usepackage[flushleft]{threeparttable}
\usepackage{hyperref}
\hypersetup{
    colorlinks=true,
    linkcolor=blue,
    filecolor=magenta,      
    urlcolor=cyan,
}
\allowdisplaybreaks
\def\baselinestretch{1.25}

\def\inclde-picture #1 by #2 (#3){%
  \vbox to #2{%
  \hrule width #1 height 0pt depth 0pt
   \vfill
   \special{picture #3}}}
\def\scaledpicture #1 by #2 (#3 scaled #4){{%
  \dimen0=#1 \dimen1=#2
  \divide\dimen0 by 1000 \multiply \dimen0 by #4
  \divide\dimen1 by 1000 \multiply \dimen1 by #4
  \inclde-picture \dimen0 by \dimen1 (#3 scaled #4)}}

\def\inclde-picture #1 by #2 (#3){%
  \vbox to #2{%
  \hrule width #1 height 0pt depth 0pt
   \vfill
   \special{picture #3}}}
\def\scaledpicture #1 by #2 (#3 scaled #4){{%
  \dimen0=#1 \dimen1=#2
  \divide\dimen0 by 1000 \multiply \dimen0 by #4
  \divide\dimen1 by 1000 \multiply \dimen1 by #4
  \inclde-picture \dimen0 by \dimen1 (#3 scaled #4)}}
\newcounter{example}

\newcounter{para}
  
\newcounter{remarkcnt}
\newcommand{\Lower}[1]{\smash{\lower 1.5ex \hbox{#1}}}
\newcommand{\HLower}[1]{\smash{\lower .03in \hbox{#1}}}
\newcommand{\LHigher}[1]{\smash{\raise .02in \hbox{#1}}}

\usepackage{natbib}
\usepackage{url}

\begin{document}

\setcounter{footnote}{1}

\vspace{.6in}
\title{\LARGE A Multi-Stage Stochastic Programming Approach to Epidemic Resource Allocation with Equity Considerations
\thanks{Cite as: Yin, X. and B\"uy\"uktahtak{\i}n\, \.I. E., 2021. A multi-stage stochastic programming approach to epidemic resource allocation with equity considerations. Health Care Management Science, pp. 1--57.}}


\author[1]{Xuecheng Yin}
\author[1]{\.I. Esra B\"uy\"uktahtak{\i}n\thanks{Corresponding author email: esratoy@njit.edu}}
\affil[1]{Department of Mechanical and Industrial Engineering, New Jersey Institute of Technology}

\maketitle

\renewcommand\baselinestretch{1.5}

\begin{abstract}
Existing compartmental models in epidemiology are limited in terms of optimizing the resource allocation to control an epidemic outbreak under disease growth uncertainty. In this study, we address this core limitation by presenting a multi-stage stochastic programming compartmental model, which integrates the uncertain disease progression and resource allocation to control an infectious disease outbreak. The proposed multi-stage stochastic program involves various disease growth scenarios and optimizes the distribution of treatment centers and resources while minimizing the total expected number of new infections and funerals. We define two new equity metrics, namely infection and capacity equity, and explicitly consider equity for allocating treatment funds and facilities over multiple time stages. We also study the multi-stage value of the stochastic solution (VSS), which demonstrates the superiority of the proposed stochastic programming model over its deterministic counterpart. We apply the proposed formulation to control the Ebola Virus Disease (EVD) in Guinea, Sierra Leone, and Liberia of West Africa to determine the optimal and fair resource-allocation strategies. Our model balances the proportion of infections over all regions, even without including the infection equity or prevalence equity constraints. Model results also show that allocating treatment resources proportional to population is sub-optimal, and enforcing such a resource allocation policy might adversely impact the total number of infections and deaths, and thus resulting in a high cost that we have to pay for the fairness. Our multi-stage stochastic epidemic-logistics model is practical and can be adapted to control other infectious diseases in meta-populations and dynamically evolving situations.
\end{abstract}

\section{Introduction}

An epidemic is the rapid spread of an infectious disease that impacts a large number of people. Epidemic diseases can disperse widely in a short time period, usually, two weeks or less, such as influenza, meningitis, and cholera, impacting populations either in a specific area, or become a pandemic affecting the lives of millions at a global scale, such as the ongoing the coronavirus pandemic. All over the world, outbreaks continue to take lives, ruin the economy, and weaken the health-care system. Unfortunately, the toll is higher in the less-developed countries because millions of people in poor regions of the world do not have the opportunity to receive sufficient treatment in case of an outbreak.

Ebola virus disease (EVD) is a prime example of a devastating epidemic. The EVD, also known as Ebola hemorrhagic fever, is a severe, often fatal illness affecting humans and other primates \citep{Ehistory}. The 2014-2016 outbreak in West Africa was the biggest Ebola outbreak in history, causing more than 28,600 cases and 11,325 deaths by the end of June 2016 \citep{Esummary}. The virus started in Guinea, then moved across countries to Sierra Leone and Liberia. The tenth outbreak of the Ebola virus disease has been ongoing in the Democratic Republic of the Congo (DRC) since August 2018. The outbreak has started from the northeast region of the country, centered in the North Kivu and Ituri provinces. More than 3000 cases have been verified by March 2020 \citep{CongoE}, and it is the country's largest-ever Ebola outbreak.

There are no specific cure, vaccine, or treatment for Ebola-infected individuals. Although multiple investigational Ebola vaccines have been developed and tested in numerous clinical trials around the world, none of them have yet been licensed to prevent the Ebola virus disease \citep{Evaccine}. Short term intervention strategies, including quarantine, isolation, contact tracing, and safe burial, have been helpful to Ebola control. Moreover, Ebola treatment centers (ETCs), which mainly isolate and treat infected individuals, play a significant role in controlling the Ebola virus disease.

The optimization problem of allocating resources to control an epidemic, such as Ebola, is an immense challenge, especially in the regions where available treatment facilities and funds are scarce. The decision-maker has to make difficult decisions to allocate limited resources to the right locations and in the right amount for slowing down the outbreak and minimize its impacts. Due to the insufficiency of intervention resources, some regions may not receive their fair share of treatment resources, compared to other regions that are also impacted by the disease. Furthermore, the EVD can spread from one individual to another through multiple mechanisms, such as through person-to person-contact or by touching the dead body infected by the EVD. The rates of disease transmission can change under various conditions and thus could be highly unpredictable.

Operations Research (OR) and mathematical modeling methods have been widely used to determine optimal resource allocation strategies to control an epidemic disease. Those approaches include simulations \citep{siettos2015modeling,ajelli2016spatiotemporal,kurahashi2015health,wells2015harnessing}, differential equations \citep{craft2005analyzing,kaplan2003analyzing}, network models \citep{berman2007location,longini2007containing,porco2004logistics,riley2006smallpox}, resource allocation analysis \citep{zaric2000methadone,tebbens2009priority,nguyen2017multiscale,shaw2010enhanced}, and stochastic compartmental models \citep{lekone2006statistical,tanner2008finding,funk2017impact, Kibisetal2019}. 

The majority of previous work focuses on analyzing the impact of different intervention strategies on disease transmission. Most of those studies consider disease growth and resource allocation problems separately in different models or enumerate each resource allocation policy in a simulation model one by one. Moreover, few studies incorporate fairness in resource allocation optimization models. The former epidemic-logistics model presented in \cite{buyuktahtakin2018new} has incorporated the logistics of treatment into a disease spread model, which foresees the disease growth over a spatial scale, and at the same time allocates limited resources to control the spread of the disease. \cite{buyuktahtakin2018new} considered the varying treatment capacity based on a limited budget. The mathematical model of \cite{buyuktahtakin2018new} was deterministic and assumed expected values for disease transmission rates. However, in reality, the disease transmission rate could be quite uncertain, changing over time and space under various scenarios. Thus, a stochastic OR model is necessary to represent the uncertainty in transmission in a more realistic way. Moreover, the majority of the OR models do not consider equity and fairness in resource allocation, resulting in solutions that may provide few or no resources to some regions impacted by the disease, especially when resources are quite limited.

The objective of this paper is to develop a multi-stage stochastic programming extension of the deterministic epidemic-logistics model of \cite{buyuktahtakin2018new} with equity considerations and present realistic insights into controlling the EVD under disease transmission uncertainty. Considering different budget levels and various tightness of the equity constraints, we analyze the optimal resource allocation strategies in a meta-population over three countries in West Africa. In our paper, the stochastic program incorporates various scenarios of disease transmission rates through person-to-person contact, thus capturing the uncertainty and variability in the infection transmission rate better compared to its deterministic counterpart. The objective function of our multi-stage stochastic programming epidemic-logistics model is to minimize the expected number of new infections and deceased individuals overall scenarios, all time periods, and all regions considered. We study the Value of the Stochastic Solution (VSS), a well-known stochastic programming measure that compares the efficiency of the deterministic and the stochastic models. Furthermore, we introduce two new equity metrics for fair resource allocation in epidemics control and analyze the impact of various budget distribution strategies on the number of infected people and deceased individuals under each of these equity metrics.

\subsection{Literature Review} \label{literature review}

The majority of mathematical models in the epidemiological literature have used simulation methods to forecast the progression of the epidemic and to study the efficacy of several interventions \citep{meltzer2014estimating,dasaklis2017emergency,pandey2014strategies, rivers2014modeling, siettos2015modeling, onal2019integrated}. Several studies have considered stochastic compartmental models to analyze different strategies for controlling epidemic diseases, such as vaccination strategies, behavioral changes that impact the interaction between different groups, and regional intervention strategies \citep{lekone2006statistical, funk2017impact}. Some other studies used simulation and network models to explore Ebola vaccination strategies \citep{nguyen2017multiscale, ball1997epidemics, shaw2010enhanced}.

Previous operations research models that study the epidemic diseases and resource allocation mainly focused on the logistics and operation management to control the disease in optimal ways \citep{zaric2001resource, buyuktahtakin2018new, ekici2013modeling, liu2019integrated}. Regarding the capacity of hospitals and logistics issues, \cite{buyuktahtakin2018new} developed a new epidemic-logistics mixed-integer programming model of the epidemic control problem. Their model considered the dynamic spread of an epidemic over multiple regions and the allocation of Ebola treatment centers and resources to control the disease simultaneously. Different than the classical epidemiological models, the transmission rate between the infected and treated compartment was not constant but instead depended on the treatment capacity and the number of infected people receiving treatment. Later, \cite{liu2019integrated} adapted the epidemics-logistics model of \cite{buyuktahtakin2018new} to study the control of the 2009 H1N1 outbreak in China and presented similar results for the H1N1 epidemic.

In the sensitivity analysis of \cite{buyuktahtakin2018new}, the disease transmission rate within the community was found to be the most critical parameter impacting infected and funerals. While the disease transmission rates are highly uncertain, relatively fewer studies in the OR community take into account the uncertain parameters for resource allocation in an effort to control the disease. Those OR models that integrate resource allocation with epidemics control use either stochastic dynamic programming (SDP), partially observable Markov decision process (POMDP) framework \citep{suen2018optimal} or two-stage stochastic programming \citep{cocsgun2018stochastic, ren2013optimal, yarmand2014optimal, tanner2008finding, long2018spatial}.

Most resource allocation models on epidemic control computed the optimal solution without considering fairness in resource allocation. Fair resource allocation has been studied in the literature, but mainly with different applications. For example, \cite{orgut2016modeling} considered a food allocation model with equitable and effective distribution of donated food under capacity constraints. \cite{davis2015improving} developed a multi-period linear optimization model for improving geographical equity in kidney allocation  while also respecting transplant system constraints and priorities. 
Moreover, \cite{lane2017equity} gave a systematic review of equity in health-care resource allocation decision-making. \cite{marsh1994equity} presented a literature review of various mathematical methods for equity measures in facility-location decision models. To our knowledge, fairness has not been studied before within the context of resource allocation for epidemic control over large spatial scales.\\

\subsection{Key Contributions and Insights} \label{literature review}
Former stochastic programming approaches on epidemic control involved a time domain with only two periods. Furthermore, there is a need for analyzing the equity and efficiency tradeoff in a mathematical programming formulation when allotting resources for controlling infectious diseases. Our approach contributes to the epidemiology and OR literature in the following ways.\\

 \noindent\textbf{Modeling Contributions.} Firstly, to the best of our knowledge, our study presents the first multi-stage stochastic programming (SP) model for infectious disease control, considering the uncertainty in the disease transmission parameter. Multi-stage SP is superior over two-stage SP models because disease transmission dynamically changes over multiple time stages. Our stochastic programming approach is also preferable to probabilistic sensitivity analysis, which considers a single scenario at a time and also to robust optimization (RO), which could provide highly conservative results by focusing on the worst-set of outcomes in a hostile environment \citep{defourny2012multistage}. Due to the temporal and spatial dimensions considered in our resource allocation model, multi-stage SP is also computationally more amenable to dynamic programming, which cannot tackle such a high-dimensional problem.

Second, we present the multi-stage VSS, which shows that the proposed stochastic programming model is superior to its deterministic counterpart.

Third, we introduce and formulate two new equity metrics and incorporate equity measures as a constraint in the mathematical formulation to balance efficiency and equity for fair resource allocation in epidemics control. To our knowledge, this study is the first one that models equity in a multi-stage stochastic programming formulation. Our multi-stage model provides an advantage of adjusting the level of equity over time with respect to evolving disease dynamics, as opposed to using a standard equity measure, which is not updated over time. Furthermore, unlike former work, we address equity in both establishing treatment centers and allocating treatment resources over metapopulations and multiple periods using mathematical optimization.

The infection equity constraint is also easier to implement than using standard equity metrics, such as the absolute difference between regional prevalence (cases per population in a region) and the overall prevalence (cases per population over all regions), because the absolute gap value using the prevalence metric could be tiny and difficult to adjust compared to the absolute gap value defined by the infection equity constraint. Furthermore, computational results imply that our model balances the proportion of infections in each region, even without including the infection equity or prevalence equity constraint.

Fourth, while we tailor our epidemics-logistics stochastic programming modeling framework for the EVD, it can be adapted to study different diseases to determine the optimal and fair resource-allocation strategies among various regions and multiple planning periods to curb the spread of an epidemic.

\noindent \textbf{Applied Contributions and Policy Insights:} Our mathematical model could be used as a decision support tool to aid policymakers in understanding disease dynamics and making the most effective decisions to fight epidemics under uncertainty. In particular, our model could be used by the stakeholders in epidemic control (e.g., governments, UN entities, non-profit organizations) to determine the optimal location and timing of ETCs opened and treatment resources allocated to minimize the total expected infections and deaths over metapopulations in multiple locations and over multiple time periods.

Our model provides significant insights into the control of the Ebola Virus Disease in West Africa that would not be possible with existing models and methods in infectious disease control. Our multi-stage stochastic program foresees various disease growth scenarios to optimize resource allocation, as opposed to solving the problem for an average scenario and myopically for one stage at a time with fixed periodic budgets, which could provide sub-optimal solutions and thus less effective resource allocation. Specifically, our study provides the following several policy insights and recommendations to decision-makers:

\begin{enumerate}[label={(\textit{\roman*})}]
\item Our analysis emphasizes that quick response, such as allocating treatment centers and resources in the early stages of the epidemic, is critical for minimizing the total number of infected individuals and deaths related to the disease.
\item The value of the stochastic solution demonstrates that the optimal timing and location of resource allocation vary with respect to the disease transmission scenario, and thus possible disease growth scenarios should be considered when planning for an epidemic instead of considering a single scenario of the expected value. 
\item Our results show that the infection level (``the number of infected people in a region'' / ``the total number of infected people'' - ``population in a region'' / ``total population'') is a key factor for resource allocation.
\item Our analysis suggests that the region with the highest infection level has the priority to receive the majority of the resources at the beginning of the time horizon to minimize the number of infections and funerals.
\item Model results also show that allocating treatment resources proportional to population is sub-optimal.
\item While equitable resource allocation is important in decision making, too much focus on the equity of resource allocation might adversely impact the total number of infections and deaths and thus resulting in a high cost that we have to pay for fairness. Therefore, decision makers are advised to be cautious about enforcing fairness when allocating resources to multiple regions.
\end{enumerate}

\section{Problem Formulation}
\label{Problem Formulation}

This section gives the formulation of a multi-stage stochastic programming model, including the compartmental model, description of the scenario tree, formulation, equity constraints, and their explanation. Model notations that will be used throughout the rest of this paper are presented in Appendix \ref{Notation}.

\subsection{Compartmental Disease Model Description}
\begin{figure}[H]
\centering
\includegraphics[width=6in]{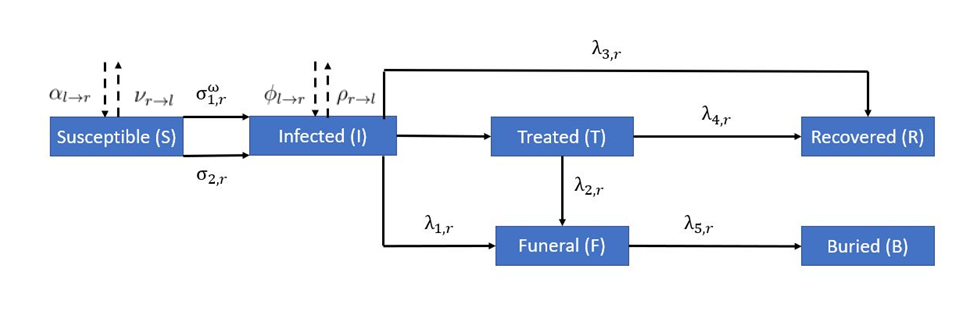}
\caption{One-Step Disease Compartmental Model}
\label{Fig34}
\end{figure}

Figure \ref{Fig34} shows the transmission dynamics of the Ebola Virus Disease (EVD) in a region $r$ of a country located in West Africa. The disease spreads among the susceptible population (S), by either person-to-person contact at a periodic rate of $\chi_{1,r}^{\omega}$ under scenario $\omega$ or through touching Ebola-related dead bodies that are not yet buried during traditional funerals at a periodic rate of $\chi_{2,r}$. Thus, susceptible individuals (S) are infected and become infected (I) with a rate of $\chi_{1,r}^{\omega}$ as a function of I and with a rate of $\chi_{2,r}$ as a function of funerals (F), who represent dead but unburied people. Without treatment,  some of the infected individuals in the compartment (I) will die and move to the funeral (F) compartment with the rate of $\lambda_{1,r}$, while some of the infected individuals will recover with a rate of $\lambda_{3,r}$, moving into the recovered compartment (R). However, the number of individuals that will be hospitalized for treatment (T) is based on the treatment capacity variable $C_{j,r}^{\omega}$, which gives the total available number of beds in the ETCs in region $r$ under scenario $\omega$ in period $j$. Thus, there is no constant transition rate from $I$ to $T$. Meanwhile, individuals who did not receive treatment will remain in the community and continue to spread the disease. In treated compartment (T), some of the individuals will recover with a periodic rate of $\lambda_{4,r}$, and a fraction of them will die with a periodic rate of $\lambda_{2,r}$. The deceased individuals in the funeral compartment are safely buried at a rate of $\lambda_{5,r}$, moving into the buried compartment (B). Thus, we assume that every death (F) leads to a safe burial (B). In order to describe the migration of susceptible and infected individuals within a given country, we define $(\alpha_{l \rightarrow r}, \upsilon_{r \rightarrow l})$ as the rates of migration of susceptible individuals into and out of region $r$, respectively, and $(\psi_{l \rightarrow r}, \rho_{r \rightarrow l})$ as the rates of migration of infected individuals into and out of region $r$, respectively. The multi-stage stochastic programming epidemic-logistics model is defined in detail in the next section. 

The latent period for the EVD is highly variable, changing from 2 to 21 days \citep{WHO2020}. In our model, we assume that each time stage represents two weeks, in which an infected but asymptomatic individual can become symptomatic and infect others. For this reason, and to avoid computational complexity, we do not include an explicit latent compartment in the model; instead, we fit those individuals within the infected compartment. Similarly, the Ebola modeling literature focusing on logistics usually omit the latent period to avoid further computational complexity (see, e.g., \cite{buyuktahtakin2018new}, \cite{long2018spatial}).

\subsection{Uncertainty Representation and Multi-period Scenario Tree Generation Scheme}
It is beyond the scope of this work to introduce a new methodology for multi-period scenario tree generation; we refer the reader to \cite{heitsch2009scenario}, \cite{leovey2015quasi}, and \cite{pflug2015dynamic} for different approaches to generate scenario trees. To generate the scenario tree for our case, we follow a similar approach presented in the study of \cite{alonso2018risk}. Here, we focus on the most uncertain parameter: the community transmission rate based on former research stating that transmission rates impact the infections and deaths the most among all different input parameters based on sensitivity analysis \citep{buyuktahtakin2018new}.

We model the future uncertainties regarding the progression of the disease by a discrete set of scenarios, denoted $\omega \in \Omega$. Each scenario has a probability, $p^{\omega}$, where $\sum_{\omega \in \Omega}p^{\omega}=1$. We assume that the uncertain community transmission rate follows a normal distribution. The data regarding the distribution of the community transmission rate parameter is not available. Thus, we use the lower and upper bounds on the transmission rate in the community based on the data gathered from literature (Table \ref{UncertainRate}) to generate the normal distribution function for the transmission rate parameter at time zero. The upper and lower bounds, thus the distribution functions for the uncertain parameter, are specified for each country and are different at each node of the scenario tree. Accordingly, the mean $\mu_r^n$ is defined for each region $r \in R$ and node $n \in N$. The lower bound and upper bound is considered as the value of 0.001- and 0.999-quantiles of the normal distribution, respectively. The standard deviation $\sigma_r$ is defined according to a normal distribution using the initial lower and upper bounds provided for each region $r \in R$. Also, we use $Q_h$ to represent the value of the $h$-quantile of the normal distribution.

\begin{table}[H]
  \centering
  \caption{The range (lower and upper bounds), mean, and standard deviation of community transmission rate in each country. Data is gathered from \cite{althaus2014estimating} and \cite{towers2014temporal}.}
    \begin{tabular}{ccccccccc}
    \hline
    Region & Rate Range & Mean & Standard Deviation \\
    \hline
     Guinea & [0.24, 0.84] & 0.54 & 0.10 \\
     Sierra Leone & [0.24, 0.88] & 0.66 & 0.07 \\
     Liberia & [0.24, 0.64] & 0.44 & 0.07 \\
    \hline
    \end{tabular}%
  \label{UncertainRate}%
\end{table}%

\begin{figure}[H]
\centering
\includegraphics[width=6in]{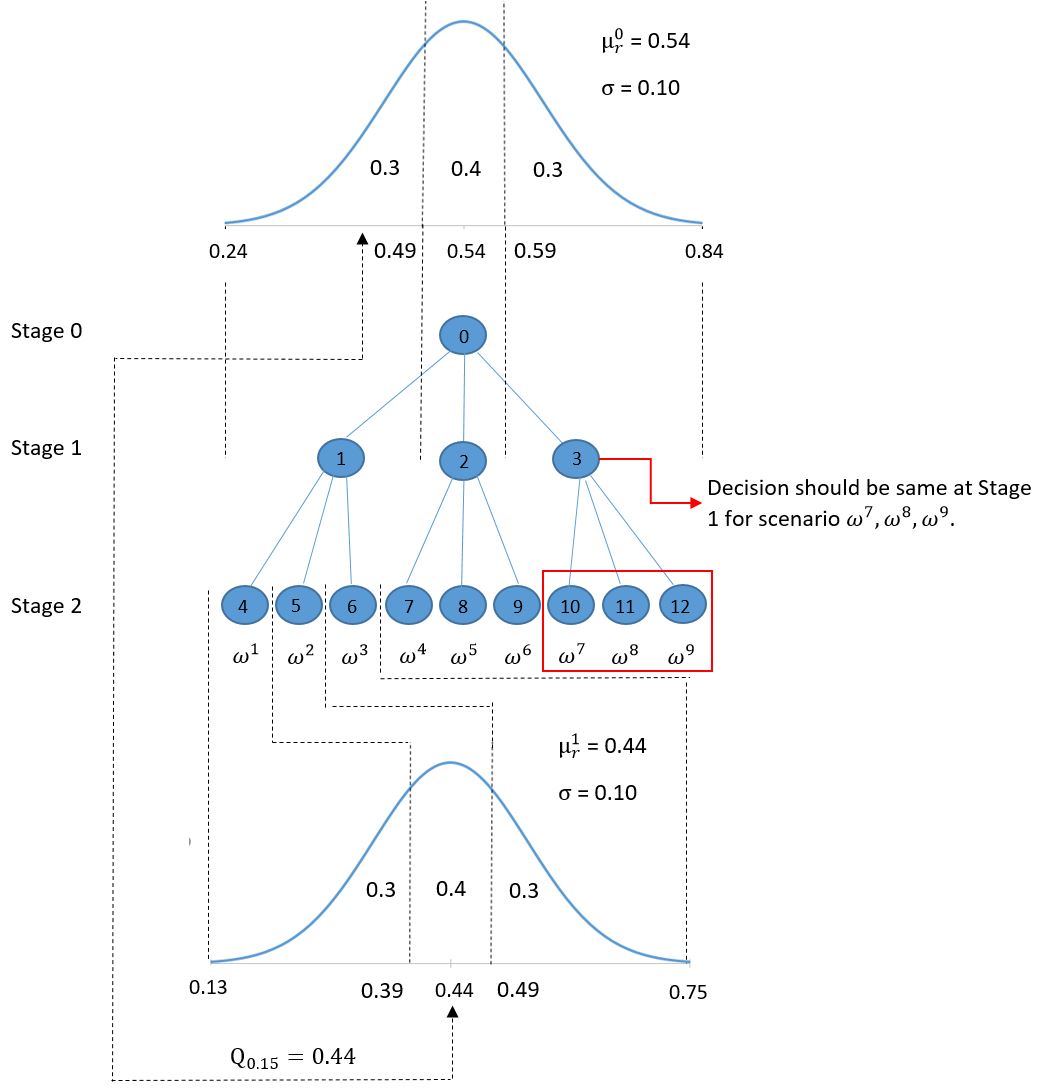}
\caption{Scenario tree generation example for Guinea, where each circle, denoted by $n$, $n:=\left\{0,\ldots, 12\right\}$, represents a node of the scenario tree.}
\label{Scheme}
\end{figure}

As shown in an example scenario tree in Figure \ref{Scheme}, a particular scenario could give the community transmission rates ($\chi_{1,r}^{\omega}$) into the future for the next two stages in all considered regions. In our model, we consider three realizations for each node of the scenario tree, namely as low, medium, and high. The low and high realizations have a probability of $0.3$, and the medium realization has a probability of $0.4$. Each path from the root node to the leaf node of the scenario tree represents a scenario $\omega$. In the example shown in Figure \ref{Scheme}, we have two stages, and thus $3^2=9$ scenarios.  In addition, two scenarios are inseparable at stage $j$ if they share the same scenario path up to that stage. This implication is modeled using non-anticipativity constraints, as described in Appendix \ref{Non-Anticipativity}. For example, for scenarios ${\omega}^1$ to ${\omega}^9$, the decision at node 0 should be the same as we do not know the values of the uncertain parameters at stage 0. Similarly, for scenarios ${\omega}^1$ to ${\omega}^3$, the decision at node 1 should be the same because these scenarios cannot be differentiated at stage 1 due to uncertainty.

The probability of a scenario $\omega$, $p^{\omega}$, is calculated as the multiplication of probabilities on the scenario path. For example, the probability of scenario ${\omega}^1$, which corresponds to a low realization in the first and second stages, is $0.09$, while the probability of scenario ${\omega}^9$, which corresponds to a medium realization in the first stage and a high realization in the second stage, is $0.12$.

For each node $n \in N$ in the scenario tree, the low  realization value of the random variable $\xi_{r}^n$ is given by the value of the 0.15-quantile ($\mu_{r,low}^n=E(\xi_{r}^n\vert Q_{0.001}\leq \xi_{r}^n \leq Q_{0.30})=Q_{0.15}$), the medium realization is given by the value of the 0.50-quantile  ($\mu_{r,medium}^n=E(\xi_{r}^n\vert Q_{0.30}\leq \xi_{r}^n \leq Q_{0.70})=Q_{0.50}$), and the high realization is equal to the value of the 0.85-quantile of the normal distribution ($\mu_{r,high}^n=E(\xi_{r}^n\vert Q_{0.70}\leq \xi_{r}^n \leq Q_{0.999})=Q_{0.85}$). In our example, at node $0$ the normal distribution of the uncertain community transmission rate parameter in Guinea has $\mu_r^0=0.54$ and $\sigma_r^0=0.10$. The low, medium, and high realizations of the uncertain parameter at nodes 0 and 1 are given in Table \ref{Q} below.
\begin{table}[H]
  \caption{The  0.15-,  0.50-,  0.85-quantiles of the normal distribution of the random variable $\xi_{r}^n$ at nodes 0 and 1 of the scenario tree in Figure \ref{Scheme}.}
  \centering
    \begin{tabular}{ccccccccc}
		 \hline
		& Low & Medium & High\\
    \hline
   node 0: & $Q_{0.15}$=0.44 & $Q_{0.50}$=0.54 & $Q_{0.85}$=0.64\\
	 node 1:& $Q_{0.15}$=0.26 & $Q_{0.50}$=0.44 & $Q_{0.85}$=0.62\\
    \hline
    \end{tabular}%
  \label{Q}%
\end{table}%

The normal distribution of community transmission rate associated with nodes 1, 2, and 3 at stage 1 have a mean of $Q_{0.15}=0.44$, $Q_{0.50}=0.54$, and $Q_{0.85}=0.64$, respectively.  While scenarios ${\omega}^1$, ${\omega}^2$, and ${\omega}^3$ at stage 1 has a single realization value of $0.54$ for the random parameter at node 1, the realizations of scenarios ${\omega}^1$, ${\omega}^2$, and ${\omega}^3$ at stage 2 correspond to nodes 4, 5, and 6, with a mean of $Q_{0.15}=0.26$, $Q_{0.50}=0.44$, and $Q_{0.85}=0.62$, respectively.

\subsection{Model Features and Assumptions}
In this study, we have considered six regions, each consisting of multiple districts, in the three countries most affected by the 2014-16 EVD. West Africa is poor and the budget for the Ebola treatment comes from an international consortium of partners, including governments, international financial Institutions, regional organizations, and private foundations \citep{UN2016}. Those funding is either directly provided to the affected governments or the United Nations (UN) entities. In this paper, we took the perspective of the UN entities, such as the World Health Organization (WHO), where the total funding is collected centrally and allocated among those three countries to optimize the use of treatment resources and the donated funding.

The actual capacity of ETCs varies from 20 to 200 operational beds \citep{WHO2020b}; however, we used 50 and 100-bed ETCs in our model to reduce the computational complexity. It is essential to differentiate the small and large ETCs in the model because each ETC type has a different setup cost, which impacts the optimal allocation of resources. We assume that each Ebola patient will receive the same treatment in either a large or small capacity ETC. The treatment capacity parameter is cumulative and only reflects total ETC beds.

Furthermore, the cost of burying dead bodies safely is shown to be minor compared to the ETC and treatment cost \citep{WHO2014, buyuktahtakin2018new}. In addition, changing the burial rate into a variable that is optimized in the model would have complicated the model considerably, and so we only focus on adjusting the variable values of treatment resources. Thus, we assume that the burial rate is constant, and burials and treatment are operated separately using different budgets.


\subsection{Model Formulation}
Following the convention of \cite{buyuktahtakin2018new}, the multi-stage stochastic programming epidemic-logistics model \eqref{Ebola:1} can be formulated as follows:
\begin{subequations}
\label{Ebola:1}
\begin{eqnarray}
\min   && \sum\limits_{j\in J-1}\sum\limits_{r\in R}\sum\limits_{\omega\in \Omega}p^{\omega}((I_{j+1,r}^{\omega}-I_{j,r}^{\omega})+F_{j+1,r}^{\omega})  \label{objective1-ex:1}\\
\textrm{s.t.} &&
        S_{0,r}^{\omega}=\pi_{r}, \quad I_{0,r}^{\omega}=\varpi_{r}, \quad T_{0,r}^{\omega}=\theta_{r}, \quad R_{0,r}^{\omega}=\vartheta_{r}, \nonumber \\  
       &&   F_{0,r}^{\omega}=\upsilon_{r}, \quad B_{0,r}^{\omega}=\tau_{r}, \quad C_{0,r}^{\omega}=\zeta_{r}, \quad r\in R, \forall\omega\in \Omega   \label{cons0-ex:1} \\
       &&   S_{(j+1),r}^{\omega}=S_{j,r}^{\omega}+\hat{S}_{j,r}^{\omega}-\widetilde{S}_{j,r}^{\omega}-\chi_{1,r}^{\omega}I_{j,r}^{\omega}-\chi_{2,r}F_{j,r}^{\omega}, \nonumber \\
       &&   j\in J\setminus \lbrace \overline{J} \rbrace, r\in R, \forall\omega\in \Omega, \label{cons2-ex:1}\\
       &&   I_{(j+1),r}^{\omega}=I_{j,r}^{\omega}+\hat{I}_{j,r}^{\omega}-\widetilde{I}_{j,r}^{\omega}+\chi_{1,r}^{\omega}I_{j,r}^{\omega}+\chi_{2,r}F_{j,r}^{\omega}-(\lambda_{1,r}+\lambda_{3,r})I_{j,r}^{\omega}-\overline{I}_{j,r}^{\omega}, \nonumber \\
       &&   j\in J\setminus \lbrace \overline{J} \rbrace, r\in R, \forall\omega\in \Omega, \label{cons2-ex:2}\\
       &&   T_{(j+1),r}^{\omega}=T_{j,r}^{\omega}+\overline{I}_{j,r}^{\omega}-(\lambda_{2,r}+\lambda_{4,r})T_{j,r}^{\omega}, \nonumber \\
       &&   j\in J\setminus \lbrace \overline{J} \rbrace, r\in R, \forall\omega\in \Omega, \label{cons2-ex:3}\\
       &&   R_{(j+1),r}^{\omega}=R_{j,r}^{\omega}+\lambda_{4,r}T_{j,r}^{\omega}+\lambda_{3,r}I_{j,r}^{\omega}, \nonumber \\
       &&   j\in J\setminus \lbrace \overline{J} \rbrace, r\in R, \forall\omega\in \Omega, \label{cons2-ex:4}\\
       &&   F_{(j+1),r}^{\omega}=F_{j,r}^{\omega}+\lambda_{1,r}I_{j,r}^{\omega}+\lambda_{2,r}T_{j,r}^{\omega}-\lambda_{5,r}F_{j,r}^{\omega}, \nonumber \\
       &&   j\in J\setminus \lbrace \overline{J} \rbrace, r\in R, \forall\omega\in \Omega, \label{cons2-ex:5}\\
       &&   B_{(j+1),r}^{\omega}=B_{j,r}^{\omega}+\lambda_{5,r}F_{j,r}^{\omega}, \qquad j\in J\setminus \lbrace \overline{J} \rbrace, r\in R, \forall\omega\in \Omega, \label{cons2-ex:6}\\
       &&   \hat{S}_{j,r}^{\omega}=\sum\limits_{l\in M_{r}}\alpha_{l \rightarrow r}S_{j,l}^{\omega}, \qquad j\in J, r\in R, \forall \omega\in \Omega, \label{cons1-ex:1}\\
       &&   \hat{I}_{j,r}^{\omega}=\sum\limits_{l\in M_{r}}\phi_{l \rightarrow r}I_{j,l}^{\omega}, \qquad j\in J, r\in R, \forall\omega\in \Omega, \label{cons1-ex:2}\\
       &&   \widetilde{S}_{j,r}^{\omega}=\sum\limits_{l\in M_{r}}\nu_{r \rightarrow l}S_{j,r}^{\omega}, \qquad j\in J, r\in R,\forall\omega\in \Omega, \label{cons1-ex:3}\\
       &&   \widetilde{I}_{j,r}^{\omega}=\sum\limits_{l\in M_{r}}\rho_{r \rightarrow l}I_{j,r}^{\omega}, \qquad j\in J, r\in R,\forall\omega\in \Omega, \label{cons1-ex:4}\\
       &&   \sum\limits_{r\in R}(\sum\limits_{j\in J\setminus \lbrace0,j\rbrace}\sum\limits_{a\in A}g_{aj,r}y_{aj,r}^{\omega}+\sum\limits_{j\in J}b_{1j,r}T_{j,r}^{\omega})\leq\Delta \quad \forall\omega\in \Omega, \label{cons3-ex:1}\\
       &&   C_{j,r}^{\omega}=\sum\limits_{m=1}^{j}\sum\limits_{a\in A}k_{a}y_{amj,r}^{\omega}+C_{0,r}, \qquad j\in J\setminus \overline{J}, r\in R,\forall\omega\in \Omega, \label{cons3-ex:2}\\
       &&   \overline{I}_{j,r}^{\omega}=min\lbrace I_{j,r}^{\omega}, C_{j,r}^{\omega}-T_{j,r}^{\omega}\rbrace, \qquad j\in J\setminus \overline{J}, r\in R,\forall\omega\in \Omega, \label{cons3-ex:3}\\
       &&   S_{j,r}^{\omega} \quad I_{j,r}^{\omega} \quad T_{j,r}^{\omega} \quad R_{j,r}^{\omega} \quad F_{j,r}^{\omega} \quad B_{j,r}^{\omega} \quad \overline{I}_{j,r}^{\omega} \geq 0, \qquad j\in J, r\in R, \forall\omega\in \Omega, \label{cons4-ex:1}\\
       &&   y_{aj,r}^{\omega}\in \lbrace 0, 1, 2, \ldots \rbrace; \quad y_{aj,r}^{\omega}\leq I_{j,r}^{\omega}, \qquad a\in A, j\in J\setminus \lbrace \overline{J} \rbrace, r\in R, \forall\omega\in \Omega, \label{cons4-ex:2}\\
       &&   y_{at(n),r}^{\omega}-y_{an,r}=0, \quad \overline{I}_{t(n),r}^{\omega}-\overline{I}_{n,r}=0, \quad C_{t(n),r}^{\omega}-C_{n,r}=0, \nonumber \\
       &&   a\in A, \forall\omega \in \beta(n), \forall n \in N, \label{cons5-ex:1}
\end{eqnarray}
\end{subequations}

The objective function \eqref{objective1-ex:1} minimizes the total expected number of newly infected individuals plus funerals over all scenarios, in all regions throughout the planning horizon. Constraints \eqref{cons0-ex:1} represent the number of individuals in susceptible, infected, treated, recovered, funeral, and buried compartments and the total ETC capacity, respectively, in each region $r$ at the beginning of the planning horizon. Equations \eqref{cons2-ex:1}--\eqref{cons2-ex:6} represent the dynamics of the population in each disease compartment, as shown in Figure \ref{Fig34}. Specifically, constraint \eqref{cons2-ex:1} implies that the number of susceptible individuals in region $r$ at the end of period $j+1$ under scenario $\omega$ is equal to the number of susceptible individuals from the previous year plus the number of susceptible individuals who immigrate into region $r$ minus the number of susceptible individuals who emigrate from region $r$ and minus the number of newly infected individuals at the end of period $j$ under scenario $\omega$. Constraint \eqref{cons2-ex:2} gives the number of infected individuals at the end of period $j+1$ in region $r$ under scenario $\omega$, which is equal to the number of infected individuals from the previous year plus immigrated infected individuals minus emigrated infected individuals, plus newly infected individuals and minus individuals who recovered, died, or were accepted for treatment at the end of period $j$ under scenario $\omega$. Constraint \eqref{cons2-ex:3} describes the total number of treated individuals in region $r$ at the end of time period $j+1$ under scenario $\omega$, which is equal to the number of treated individuals at the end of period $j$ plus infected individuals who accepted treatment based on the availability of beds minus treated individuals who died or recovered. Constraint \eqref{cons2-ex:4} ensures that the cumulative number of recovered individuals in region $r$ at the end of the period $j+1$ under scenario $\omega$ is equal to the number of recovered individuals from the previous year plus newly recovered individuals. Constraint \eqref{cons2-ex:5} defines the total number of unburied funerals in region $r$ at the end of time period $j+1$ under scenario $\omega$, which is equal to the infected and treated individuals who moved to the funeral compartment minus the buried dead bodies. Constraint \eqref{cons2-ex:6} gives the cumulative number of buried dead bodies at the end of the period $j$ under scenario $\omega$. Constraints \eqref{cons1-ex:1}--\eqref{cons1-ex:4} present the number of immigrated and emigrated individuals in susceptible and infected compartments. Specifically, constraints \eqref{cons1-ex:1} and \eqref{cons1-ex:2} give the number of susceptible and infected individuals who immigrated into region $r$ from region $l \in M_r$ under scenario $\omega$. Constraints \eqref{cons1-ex:3} and \eqref{cons1-ex:4} represent the number of susceptible and infected individuals, who emigrated from region $r$ into neighboring region $l \in M_r$ under scenario $\omega$. Constraints \eqref{cons3-ex:1}--\eqref{cons3-ex:3} represent the restrictions regarding logistics and operation management. Specifically, constraint \eqref{cons3-ex:1} denotes the budget limitation on the sum of the fixed costs of opening ETCs and the variable cost of treating infected individuals over all regions $r$ in all periods $j$ under scenario $\omega$. Constraint  \eqref{cons3-ex:2} shows the total capacity in region $r$ at the end of period $j$ under scenario $\omega$. Constraint \eqref{cons3-ex:3} ensures that the number of hospitalized individuals is limited by the number of available beds in ETCs in region $r$. In particular, the number of hospitalized individuals ($\overline{I}$) is equal to the minimum of the number of infected individuals and the capacity available at established ETCs after considering currently hospitalized individuals in ETCs. Constraints \eqref{cons4-ex:1} present non-negativity restrictions on the number of susceptible, infected, treated, funeral, buried, and recovered individuals, respectively, under scenario $\omega$. Constraints \eqref{cons4-ex:2} denote the integer requirements on the number of type-$n$ ETCs to be opened in region $r$ at the end of period $j$ under scenario $\omega$. In addition, if the number of infected individuals is less than 1 in a region $r$, the value of the integer variable corresponding to opening an $n$-bed ETC is forced to be zero, and thus no ETC will be opened in that region. Constraints \eqref{cons5-ex:1} represent nonanticipativity restrictions, which state that if two scenarios share the same path up to stage $j$, the corresponding decisions should be the same, as described in Appendix \ref{Non-Anticipativity}.

\subsection{Equity of ETC and treatment resources distribution} \label{Equity Constraints}

Equitable resource allocation has long been studied in health-care resource allocation decision making \citep{lane2017equity}. Some examples include equity in facility location \citep{marsh1994equity, ares2016column}, organ allocation for kidney transplantation \citep{su2006recipient, bertsimas2013fairness}, vaccine coverage \citep{enayati2020optimal}, and health-care fleet management \citep{mccoy2014using}.

In the health-care sector, an equity metric compares two or more populations based on the service or utility the health system provides to the different populations. The comparison of various populations could be based on the health status, distribution of resources, expenditures, utilization, and access \citep{goddard2001equity, culyer1993equity}.

While it is essential to clearly define equity to be used for fair resource allocation, there is no universal consensus on the definition and measurement of equity in public health decision making \citep{stone2002policy}. \cite{lane2017equity} find a large disparity in the description of equity in health care resource allocation based on their review of the related literature.

Among numerous definitions of equity, \cite{young1995equity} defines three equity concepts on resource allocation: parity (claimants should be treated equally), proportionality (goods should be divided in proportion to differences among claimants), and priority (the person with the greatest claim to the good should get it). \cite{savas1978equity} describes equity as fairness, impartiality, or equality of service. \cite{culyer2001equity} discusses utilitarian principles dictating that resources should be allocated in such a way as to maximize the overall health and wellbeing of a society, and egalitarian principles dictating that all people are equal and that inequalities between groups should be removed. \cite{mccoy2014using} use utilitarian, proportionally fair, and egalitarian principals to incorporate equity into optimal resource allocations.

\cite{marsh1994equity} present a list of 20 equity measures within the context of facility location. Among the most commonly-used equity measures are the sum of absolute deviations (SAD), the mean absolute deviation (MAD), the minimum effect (ME), and the Gini coefficient (GC). \cite{love2020methods}  categorize methods used to define equity measures into five: 1) gap measures, regression-based measures, Lorenz and concentration curves, measures incorporating inequality aversion, and health-related social welfare. The equity measures defined by absolute and relative gaps are commonly used by international agencies, such as the WHO, to distribute resources, such as vaccines and medical treatment, between population groups in low- and middle-income countries \citep{casey2017state}.

Equitable resource allocation has also been studied considering the tradeoff between the efficiency and equity in resource allocation for infectious diseases, such as HIV and influenza (e.g., \cite{mbah2011resource}, \cite{zaric2007little}, \cite{kaplan2002allocating}, \cite{enayati2020optimal}). For example, \cite{earnshaw2007linear} develop a linear programming planning tool to help policymakers understand the effectiveness of different allocations of HIV prevention funds under fairness constraints. \cite{enayati2020optimal} propose an equity constraint in a mathematical program to help public health authorities consider fairness when making vaccine distribution decisions. In a food allocation problem, \cite{orgut2016modeling} present a deterministic linear programming model to optimize the allocation of donated food, considering objectives of both equity and effectiveness.

Similar to these works, we will follow an approach that would balance the efficiency and equity in epidemics resource allocation. Specifically, we will focus on equity over meta-populations and multiple spatial dimensions. We define our equity measures within the context of proportionality and priority, as described in \cite{young1995equity}. Our formulations of equity are gap-based, combining absolute and relative gaps. Our approach is seeking a balance between utilitarian and egalitarian objectives studied in 
\cite{culyer1993equity} and \cite{mccoy2014using} by determining a resource allocation strategy that will minimize total infections and deaths but at the same time incorporates equality dimensions as a constraint. Unlike former work, we address equity in the resource allocation for both treatment centers and treatment resources using mathematical optimization.

Our definition of equity is similar to the descriptions of \cite{mbah2011resource}, who defines social equity as the equal opportunity for infected individuals to access treatment, \cite{marsh1994equity}, who define equity within the context of facility location, and  \cite{orgut2016modeling} who study equity in the fair allocation of food. Specifically, we define equity as the case where each region and country receives its fair share of the ETCs and medical treatment resources during an epidemic outbreak.

The majority of studies on fair resource allocation define the equity as a one-period metric, which does not change over time. In our multi-stage stochastic programming model, the equity standard is adjusted over time with respect to the changing disease dynamics throughout the planning horizon, increasing the efficiency of the resource allocation. To the best of our knowledge, our study is the first to model the fair resource allocation using a multi-stage stochastic programming model, which captures disease dynamics over multiple time periods.

\subsubsection{Infection Equity Constraint}

In the first formulation, we will address the objective of equity by limiting the absolute deviation between a region’s relative number of infections and its relative population with respect to all regions, while effectiveness corresponds to minimizing the expected number of infections and deaths. In this equity measure, namely infection equity constraint, we consider priority concerning the proportions of infections and enforce resource allocation to limit the proportion of infections with respect to the population for each region. The infection equity constraint is given as follows:

\begin{equation} 
\vert\frac{\sum\limits_{j \in J}\sum\limits_{\omega \in W}p^{\omega} I_{j,r}^{\omega}}{\sum\limits_{j \in J}\sum\limits_{r \in R}\sum\limits_{\omega \in W}p^\omega I_{j,r}^{\omega}} - \frac{u_r}{\sum\limits_{r \in R}u_r} \vert \leq k \label{cons3-ex:4} 
\end{equation} 		
The infection equity constraint \eqref{cons3-ex:4} gives a bound on the total number of infections in each region relative to the total infections in all regions. Specifically, constraint \eqref{cons3-ex:4} implies that the absolute value of the number of infected individuals in region $r$ divided by the total number of infected individuals over all regions minus the ratio of the population of region $r$, $u_r$, over the total population over all regions should be less than or equal to a specific value $k$. \\

Because the EVD case fatality rate is high [50\% on average \citep{WHO2020}] and the EVD is highly contagious, having the lowest infections system-wide will lead to the lowest mortality for the EVD. Thus, we consider the number of infections instead of deaths as the main parameter for resource allocation in our equity metric. The number of infections in constraint (2) could also be adjusted to the number of fatalities.

\subsubsection{Capacity Equity Constraint}

In the second formulation, we will formulate equity by limiting the absolute deviation between the proportion of treatment capacity established in a region and proportion of the population in a region relative to all regions while again, effectiveness corresponds to minimizing expected deaths and infections. The capacity equity constraint enforces allocating resources considering the proportionality based on the relative population and is formulated as follows:

\begin{equation} 
\vert\frac{\sum\limits_{j \in J}\sum\limits_{\omega \in W}p^{\omega} C_{j,r}^{\omega}}{\sum\limits_{j \in J}\sum\limits_{r \in R}\sum\limits_{\omega \in W}p^\omega C_{j,r}^{\omega}} - \frac{u_r}{\sum\limits_{r \in R}u_r} \vert \leq k ,   \label{cons3-ex:5}
\end{equation} 

Similarly, we define the capacity equity constraint \eqref{cons3-ex:5} to bound the absolute value of the difference between the proportion of the capacity at region $r$ over the total capacity with the proportion of the population at region $r$ over the total population with a predefined parameter $k$.

\subsubsection{Prevalence Equity Constraint}

We also study a widely-used equity metric, known as prevalence \citep{lasry2008s4hara, kedziora2019optimal}. Here, we define the prevalence equity constraint to limit the absolute difference between the regional prevalence (cases per population in a region) and the country prevalence (cases per population over all regions) by the parameter $k$, and formulate it as follows:

\begin{equation} 
\vert\frac{\sum\limits_{j \in J}\sum\limits_{\omega \in W}p^{\omega} I_{j,r}^{\omega}}{u_r} - \frac{\sum\limits_{j \in J}\sum\limits_{r \in R}\sum\limits_{\omega \in W}p^\omega I_{j,r}^{\omega}}{\sum\limits_{r \in R}u_r} \vert \leq k \label{Prevalence Equity Constraint} 
\end{equation}

The prevalence equity constraint \eqref{Prevalence Equity Constraint} bounds the proportion of infections in each region relative to the proportion of infections in all regions.

\subsection{Mixed-Integer Linear Program (MIP) Model}
\label{Mixed-Integer Linear Program (MIP) Model}

In the mathematical formulation \eqref{Ebola:1}, we have two types of non-linearity. The first non-linear equation corresponds to the capacity-availability constraint \eqref{cons3-ex:3}, and the second corresponds to the equity constraints \eqref{cons3-ex:4} and \eqref{cons3-ex:5} (see Appendix \ref{Linearization} for linearization of \eqref{cons3-ex:3}, \eqref{cons3-ex:4}, and \eqref{cons3-ex:5}). The non-linear multi-stage stochastic programming epidemic–logistics model \eqref{Ebola:1} is converted into an equivalent MIP formulation by replacing the non-linear capacity availability constraint \eqref{cons3-ex:3} with constraints \eqref{lin1-ex2}, \eqref{lin1-ex3}-\eqref{lin1-ex4} and \eqref{lin1-ex5}-\eqref{lin1-ex6}, the non-linear infection equity constraint \eqref{cons3-ex:4} with constraints \eqref{lin2-ex2} and \eqref{lin2-ex3}, and the non-linear capacity equity constraint \eqref{cons3-ex:5} with constraints \eqref{lin2-ex4} and \eqref{lin2-ex5}, as given in Appendix \ref{Linearization}.

We apply the MIP model to a case study involving the control of the 2014–2015 Ebola outbreak in the three most-affected West African countries, Guinea, Sierra Leone, and Liberia. The details of the 2014–2015 Ebola outbreak data used as an input into the mathematical model, including population and migration data, resource cost data, and epidemiological data are presented in Appendix \ref{Ebola Case Study Data}. 

The MIP model is solved using CPLEX 12.7 on a desktop computer running with Intel i7 CPU and 64.0 GB of memory. A time limitation of 7,200 CPU seconds was imposed for solving the test instances without equity constraints, while the time limit is increased to 72,000 CPU seconds for the instances with equity constraints due to their computational difficulty. The multi-stage stochastic model is solved over eight stages for the base case with each stage representing a 2-week period, thus for a total of the 16-week planning horizon. Since we consider three outcomes on each branch of the scenario tree, we solve for $3^8=6561$ scenarios in the mathematical model.

\section{Results} \label{results}
In this section, we present computational results for the multi-stage stochastic MIP model presented in Section \ref{Problem Formulation} for the considered case study instance in West Africa. Our goal in this section is to provide insights into the optimal and fair resource allocation for controlling the Ebola disease outbreak under the uncertainty of disease transmission. 

\subsection{Model Validation}
In this subsection, we validate our model against the real outbreak data \citep{Outbreak} in terms of the cumulative number of infections from August 30, 2014, to December 19, 2014. The values of parameters used in the model are obtained from the literature \citep{camacho2014potential,who2014ebola,WHO}.

We fix the number of ETCs at each stage according to the number and timing of the ETCs established in reality \citep{buyuktahtakin2018new}. For instance, according to the outbreak data, one 50-bed ETC was established on September 15, 2014, in northern Liberia, and so the value of the related variable is fixed to one in stage one in the model. Once the ETCs are fixed in the model based on their opening time and the capacity throughout the planning horizon, the model is solved and validated by comparing the predicted number of infections with the real outbreak data given in the WHO database \citep{Outbreak}.

\begin{figure}[H]
\centering
\includegraphics[width=6in]{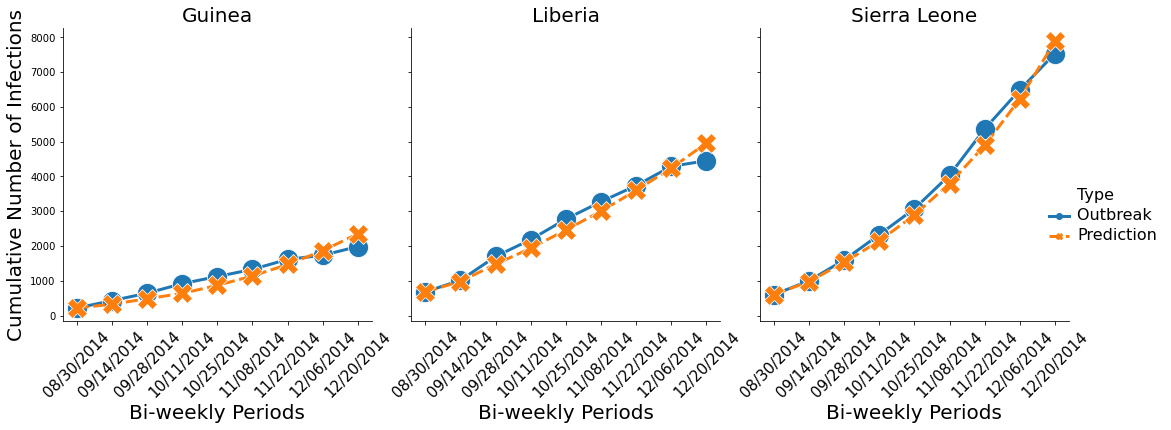}
\caption{Comparison of predicted cases with real outbreak data for cumulative infections in Guinea, Liberia, and Sierra Leone}
\label{Compare}
\end{figure}

According to the visual comparison of the predicted results and real outbreak data in Figure \ref{Compare}, our model provides a good fit for the cumulative number of infected individuals in Guinea, Sierra Leone, and Liberia during the considered time period. In addition, we apply the paired t-test to analyze the difference between the pairs of weekly predicted cases and the actual data. As shown in Table \ref{pairedttest}, all p-values are greater than 0.05, indicating that our model provides statistically similar results to the real outbreak data from August 30, 2014, to December 19, 2014.

\begin{table}[H]
  \centering
  \caption{Statistical analysis to compare bi-weekly predicted cases and real outbreak data.}
    \begin{tabular}{ccccccccc}
    \hline
     & Country & \multicolumn{2}{c}{Mean} & \qquad & \multicolumn{3}{c}{Two-tailed paired t-test} \\
     \cline{3-4}\cline{6-8}
     & & Outbreak & Predicted & & t-stat & t-critical & p-value \\
    \hline
      & Guinea & 221.0 & 266.8 & & 0.41 & 1.89 & 0.65\\
    Infections  & Sierra Leone & 866.3 & 910.1 & & 0.65 & & 0.73 \\
      & Liberia & 471.1 & 534.5 & & 0.45 & & 0.67\\
    \hline
    \end{tabular}%
  \label{pairedttest}%
\end{table}%


\subsection{The Value of Stochastic Solution (VSS)}
To demonstrate the value of using a stochastic program over a deterministic (expected value) model, we use a standard measure in stochastic programming, known as the value of stochastic solution (VSS) \citep{birge1982value}. The VSS gives the expected gain from solving a stochastic model over its deterministic counterpart, in which random parameters are replaced by their expected values. 


\subsubsection{Two-Stage VSS}

WS is the wait-and-see problem objective value, which is the expected value of using the optimal solution for each scenario. EEV is the expected result of using the solution of the deterministic model (EV), which replaces all uncertain parameters by their expected values, and RP is the optimal value of our stochastic programming model, i.e., the minimization recourse problem. Then the following inequalities are satisfied for the minimization problems \citep{madansky1960inequalities}:
$$WS \leq RP \leq EEV,$$ 
The VSS can then be formulated as follows:
$$VSS = EEV - RP$$
A large value of the VSS implies that incorporating uncertainty is important to represent the problem realistically, and the solution of the deterministic problem is not ``so good.'' On the other hand, if the VSS value is small, replacing uncertain parameters with their expected values might be a good choice. 


\subsubsection{Multi-Stage VSS}

For the multi-stage problem, the value of the stochastic solution is introduced as a chain of values $VSS_{t}$ for $t=1,\ldots,T$, where $T$ is the final period of the planning horizon \citep{escudero2007value}. In order to calculate the $VSS_{t}$, the solution up to stage $t-1$ of the associated deterministic model is fixed in the stochastic model resulting in the $EEV_t$ value, and RP value is subtracted from $EEV_t$. Consider a stochastic model, which only contains decision variables $x$ and recourse variables $y$, and let $(\hat{x_t}, \hat{y_t})$ be the optimal solution of the corresponding EV model. The $EEV_t$ can then be formulated as:
\begin{subequations}
\begin{eqnarray}
\ EEV_t: RP   &&   \text{model} \nonumber\\
\textrm{s.t.} &&
        x_1^{\omega} = \hat{x}_1 \quad \forall\omega\in \Omega, \nonumber \\
        &&   \cdots \nonumber \\
        &&   x_{t-1}^{\omega} = \hat{x}_{t-1} \quad \forall\omega\in \Omega. \nonumber  
\end{eqnarray}
\end{subequations}

The $VSS_{t}$ for each $t=1,\ldots,T$ is then given as:
$$VSS_{t} = EEV_{t} - RP$$

As an example, we calculate the $VSS_t$ for an $8$-stage problem for $t=1,\ldots,4$. Since $EEV_1 = RP$, the value of the $VSS_{1}$ is zero. We solve the model under a \$24M budget and present the results in Table \ref{4-stageVSS} below.
\begin{table}[H]
  \centering
  \caption{$VSS_{t}$ values up to 4 stages for the 8-stage problem with $EEV_{t}$ values}
    \begin{tabular}{ccccccccc}
    \hline
    $VSS_1 (RP)$ & $VSS_2$ & $VSS_3$ & $VSS_4$ \\
    \hline
    0 & 41 & 65 & 69 \\
    \hline
    \end{tabular}%
  \label{4-stageVSS}%
\end{table}%

The RP value for the $8$-stage problem is 2207 individuals. The $VSS_t$ value is increasing as the stage $t$ increases, thus a multi-stage stochastic model is needed to obtain a better result compared to the deterministic problem. We notice that under the \$24M budget level, the model allocates almost all the ETCs in the first stage. Thus, the $VSS_t$ value will not change significantly when $t\geq 3$. For varying budget cases or disease dynamics, we expect that the model will allocate ETCs in the stages following the first stage, and thus the $VSS_t$ values may become larger than the values in this instance.  The results for solving the $8$-stage model highlight the importance of using a multi-stage stochastic model for the epidemic-logistics problem over its deterministic counterpart.


\subsection{Analysis of Budget Allocation}


The columns of Table \ref{BudgetResult1} present results for each \textbf{Budget} level (\$12M, \$24M, and \$48M), each \textbf{Country} and \textbf{Region}, \textbf{Stage-1 Budget} allocated, \textbf{Total Budget} allocated,  \textbf{Stage-1 ETC (50/100)} representing the number of 50- and 100-bed ETCs allocated in the first stage of the planning horizon, and \textbf{Total ETC (50/100)} indicating the total number of 50- and 100-bed ETCs allocated throughout the planning horizon. Here, expected values of the optimal budget and the number of ETCs allocated at the first stage and throughout the planning horizon over 6561 scenarios are presented for each budget level. Correspondingly, the expected value of the total number of infections and funerals for different budget levels is presented in Figure \ref{Fig45}. The CPU time used to solve the model is $7230s$ for the \$12M budget, $7232s$ for the \$24M budget, and $7228s$ for the \$48M budget. The optimality gap for all the cases are 0.1\%.

\begin{figure}[H]
\centering
\includegraphics[width=6in]{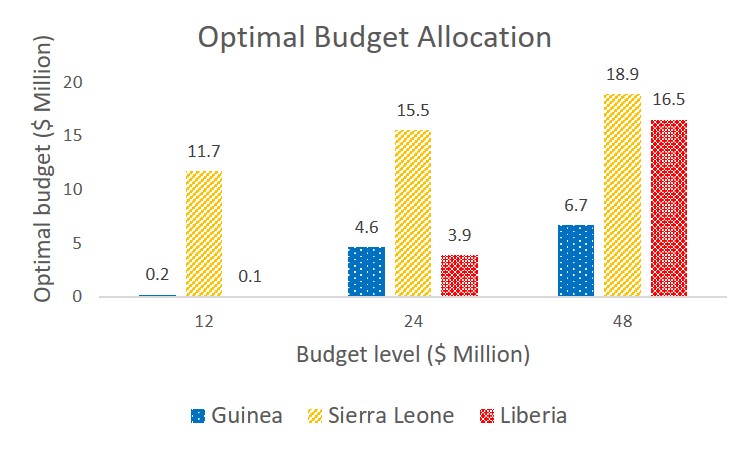}
\caption{Total budget allocation under different budget levels}
\label{Fig44}
\end{figure}

\begin{figure}[H]
\centering
\includegraphics[width=6in]{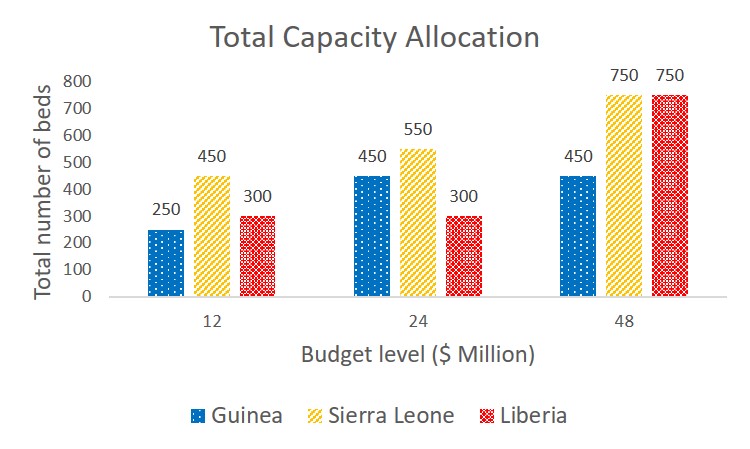}
\caption{Total capacity allocation under different budget levels}
\label{Fig42}
\end{figure}

\begin{figure}[H]
\centering
\includegraphics[width=6in]{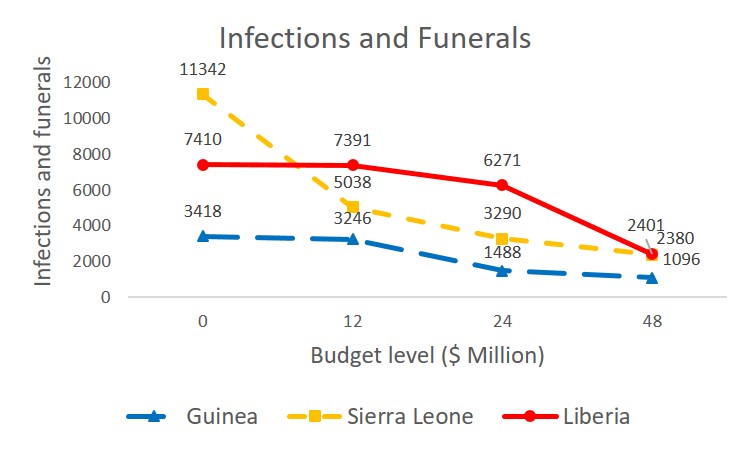}
\caption{Total number of infections and funerals under different budget levels}
\label{Fig45}
\end{figure}

The fifth column of Table \ref{BudgetResult1} and Figure \ref{Fig44} show the allocation of the total budget among three different countries. Due to the high initial number of infected individuals, Sierra Leone gets the most budget allocation under all different budget levels. Although the transmission rate of Guinea is higher than Liberia, the second highest budget goes to Liberia under the \$48M budget case because the initial state of the infection in this country is high, and thus, the allocated budget will provide a more significant impact on Liberia compared to Guinea when the budget is ample. According to the results of ETC allocation at all budget levels, most of the beds are allocated in the first period (stage-1) of the planning horizon under tight budget cases, as shown in Table \ref{BudgetResult1}. Figure \ref{Fig42} shows the total capacity allocation under different budget levels.

Figure \ref{Fig45} shows the total number of infections and funerals in those three countries under different budget levels. According to the result under the \$0M budget level, the case in which no intervention action is taken, the number of infections and funerals in Sierra Leone would be extremely large if we do not take any intervention action. As shown in Figure \ref{Fig45}, the total number of infections and funerals in all three countries, especially in Liberia and Sierra Leone, drops significantly from \$12M to \$48M budget level.

\begin{table}[H] \small
		\renewcommand\arraystretch{0.8}
\renewcommand\tabcolsep{3.0pt}
  \centering
  \caption{Budget and bed allocated under different budget levels}
    \begin{tabular}{ccrcccccc}
    \hline
		   \textbf{Budget} & \textbf{Country} & \textbf{Region} & \textbf{Stage-1 } 			& \textbf{Total}         & \textbf{Stage-1}     & \textbf{Total}  \\
       (\textbf{\$M})  &         &        & \textbf{Budget}   & \textbf{Budget}  & \textbf{ETC}   & \textbf{ETC} \\
			      &         &        & \textbf{(\$M)}   & \textbf{(\$M)}  & \textbf{(50/100)}    & \textbf{(50/100)} \\
    \hline
    \multirow{6}[1]{*}{12} & \multirow{3}[1]{*}{Guinea} & \multicolumn{1}{c}{UG} & 0.06 & 0.13  & 1/1   & 1/1 \\
          &       & \multicolumn{1}{c}{MG} & 0.01     & 0.02 & 1/0     & 1/0 \\
          &       & \multicolumn{1}{c}{LG} & 0.03     & 0.06  & 1/0     & 1/0   \\
          & Sierra Leone & \multicolumn{1}{c}{S} & 4.33   & 11.67 & 1/4   & 1/4 \\
          & \multirow{2}[0]{*}{Liberia} & \multicolumn{1}{c}{NL} & 0.04     & 0.09     & 1/1     & 1/1 \\
          &       & \multicolumn{1}{c}{SL} & 0.01     & 0.03 & 1/1     & 1/1     \\
    \textbf{Total} &       &       & \textbf{4.47}   & \textbf{11.99}  & \textbf{6/7}   & \textbf{6/7} \\
    \hline
    \multirow{6}[1]{*}{24} & \multirow{3}[1]{*}{Guinea} & \multicolumn{1}{c}{UG} & 0.72     & 1.83  & 1/1     & 1/1 \\
          &       & \multicolumn{1}{c}{MG} & 0.52     & 1.21  & 1/0 & 1/1   \\
          &       & \multicolumn{1}{c}{LG} & 0.62     & 1.53  & 1/1 & 1/1   \\
          & Sierra Leone & \multicolumn{1}{c}{S} & 5.35   & 15.50 & 1/5   & 1/5 \\
          & \multirow{2}[0]{*}{Liberia} & \multicolumn{1}{c}{NL} & 0.83   & 2.31   & 1/1   & 1/1 \\
          &       & \multicolumn{1}{c}{SL} & 0.57   & 1.60   & 1/1   & 1/1   \\
    \textbf{Total} &       &       & \textbf{8.62}  & \textbf{23.98}  & \textbf{6/9}  & \textbf{6/10} \\
    \hline
    \multirow{6}[1]{*}{48} & \multirow{3}[1]{*}{Guinea} & \multicolumn{1}{c}{UG} & 1.11 & 2.52   & 1/1   & 1/1 \\
          &       & \multicolumn{1}{c}{MG} & 0.91     & 1.91   & 1/1     & 1/1   \\
          &       & \multicolumn{1}{c}{LG} & 1.01     & 2.32   & 1/1     & 1/1 \\
          & Sierra Leone & \multicolumn{1}{c}{S} & 6.85   & 18.89 & 4/5   & 5/5 \\
          & \multirow{2}[0]{*}{Liberia} & \multicolumn{1}{c}{NL} & 3.94   & 10.42 & 3/3   & 3/3 \\
          &       & \multicolumn{1}{c}{SL} & 2.40   & 6.03  & 2/2   & 2/2  \\
    \textbf{Total} &       &       & \textbf{16.22}  & \textbf{42.09}  & \textbf{12/13}  & \textbf{13/13} \\
   \hline
    \end{tabular}%
  \label{BudgetResult1}%
\end{table}%

The results presented in this section represent the expected values over all scenarios. To perform a more detailed analysis, we picked 5 (five) out of 6561 scenarios and analyzed the corresponding results in the next section.

\subsection{Analysis of Different Scenarios}

In this section, we present results regarding the budget, and ETC allocation as well as the corresponding total number of infections and funerals for five specific scenarios under a budget level of \$24M. Those four different scenarios are defined as follows. The first scenario is the ``All Low'' case that corresponds to the low realization of the uncertain community disease transmission rate from stages 1 to 8, the second scenario is the ``All Medium'' case that corresponds to the medium realization of the uncertain community disease transmission rate from stages 1 to 8, the third scenario is the ``All High'' case that represents the high realization of the community disease transmission rate from stages 1 to 8, the fourth scenario is the ``Low-High'' case that stands for the low realization of the disease transmission rate from stages 1 to 4 followed by its high realization from stages 5 to 8, and the fifth scenario is the ``High-Low'' case that represents the high realization of the community disease transmission rate from stages 1 to 4 followed by a low transmission rate from stages 5 to 8. According to the results, we divided scenarios into two groups except for the ``All Medium'' case; the first one is called the better group, including ``All Low'' and ``Low-High'' cases, on the other hand, the second group is called the worse group, encompassing ``All High" and ``High-Low'' cases. Similar to Table \ref{BudgetResult1}, Table \ref{ScenarioResult1} presents results for each \textbf{Scenario} defined above under the \$24M budget level.

\begin{figure}[H]
\centering
\includegraphics[width=6in]{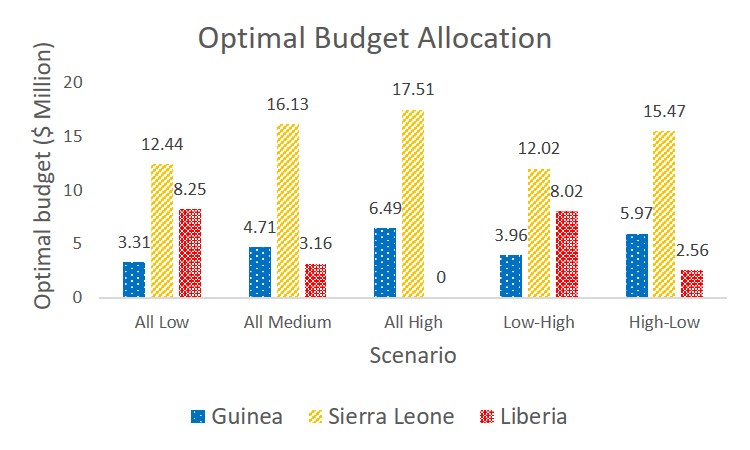}
\caption{Total budget allocation under different scenarios}
\label{Fig47}
\end{figure}

The first-stage budget allocation is presented in the fourth column of  Table \ref{ScenarioResult1}, while the total budget is presented in both the fifth column of Table \ref{ScenarioResult1} and Figure \ref{Fig47}. In terms of bed allocation, all the regions have the same number of bed allocation for stage-1 and for the total stages under all scenarios. This result implies that it is optimal to open treatment centers early in all the locations, in particular, in the initial stages.

\begin{figure}[H]
\centering
\includegraphics[width=6in]{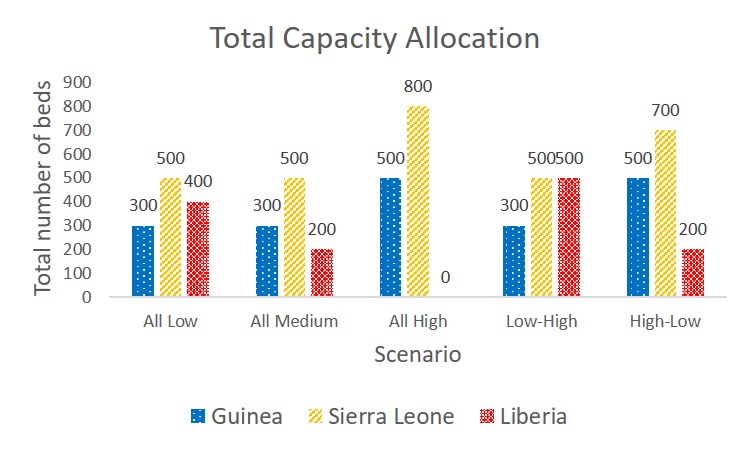}
\caption{Total capacity allocation under different scenarios}
\label{Fig46}
\end{figure}

\begin{figure}[H]
\centering
\includegraphics[width=6in]{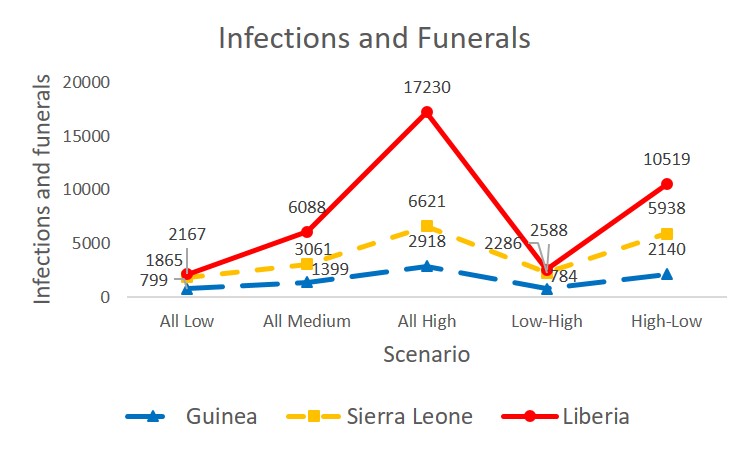}
\caption{Total number of new infections and funerals under different scenarios}
\label{Fig48}
\end{figure}

Figure \ref{Fig46} represents the total capacity allocation under different scenarios. According to the results, the total capacity allocated under the worse scenario group is higher than the capacity allocated for the better group. This result implies that under the worse scenario group, more budget is allocated to build new Ebola treatment centers. Also, as shown in Figure \ref{Fig48}, the total number of new infections and funerals under the ``High-Low'' case is much higher than the corresponding number under the ``Low-High'' case. Thus, a scenario where the disease starts with a low transmission rate and then progresses fast is better than a scenario in which the disease progression is fast and then slows down. This may be because diseases that initially progress less aggressively give us more time to get prepared, establish the ETCs and treatment resources, and thus reduce the number of infections immediately.

\begin{table}[H] \small
		\renewcommand\arraystretch{0.7}
\renewcommand\tabcolsep{3.0pt}
  \centering
  \caption{Budget and bed allocated under different scenarios}
    \begin{tabular}{ccrcccccc}
    \hline
		   \textbf{Scenario} & \textbf{Country} & \textbf{Region} & \textbf{Stage-1 } 			& \textbf{Total}     & \textbf{Total} \\
       (\textbf{\$M})  &         &        & \textbf{Budget}   & \textbf{Budget}   & \textbf{Bed} \\
			      &         &        & \textbf{(\$M)}   & \textbf{(\$M)}    & \textbf{(50/100)} \\
    \hline
    \multirow{6}[1]{*}{All Low} & \multirow{3}[1]{*}{Guinea} & \multicolumn{1}{c}{UG} & 0.60    & 1.15 & 2/0 \\
          &       & \multicolumn{1}{c}{MG} & 0.60   & 1.02 & 2/0 \\
          &       & \multicolumn{1}{c}{LG} & 0.60     & 1.13 & 2/0 \\
          & Sierra Leone & \multicolumn{1}{c}{S} & 5.39 & 12.44  & 0/5 \\
          & \multirow{2}[0]{*}{Liberia} & \multicolumn{1}{c}{NL} & 2.15 & 5.46 & 0/2 \\
          &       & \multicolumn{1}{c}{SL} & 1.08 & 2.79 & 0/2 \\
    \textbf{Total} &       &       & \textbf{10.41}     & \textbf{24.00}     & \textbf{6/9} \\
    \hline
    \multirow{6}[1]{*}{All Medium} & \multirow{3}[1]{*}{Guinea} & \multicolumn{1}{c}{UG} & 0.60 & 1.64 & 2/0 \\
          &       & \multicolumn{1}{c}{MG} & 0.60 & 1.43 & 2/0 \\
          &       & \multicolumn{1}{c}{LG} & 0.60 & 1.64 & 2/0 \\
          & Sierra Leone & \multicolumn{1}{c}{S} & 5.39 & 16.13  & 0/5 \\
          & \multirow{2}[0]{*}{Liberia} & \multicolumn{1}{c}{NL} & 0 & 0 & 0/0 \\
          &       & \multicolumn{1}{c}{SL} & 1.08 & 3.16 & 0/2 \\
    \textbf{Total} &       &       & \textbf{8.26}   & \textbf{24.00}  & \textbf{6/7} \\
    \hline
    \multirow{6}[1]{*}{All High} & \multirow{3}[1]{*}{Guinea} & \multicolumn{1}{c}{UG} & 1.08 & 2.75 & 0/2 \\
          &       & \multicolumn{1}{c}{MG} & 0.60 & 1.51 & 2/0 \\
          &       & \multicolumn{1}{c}{LG} & 1.08 & 2.23 & 0/2 \\
          & Sierra Leone & \multicolumn{1}{c}{S} & 7.06 & 17.51  & 2/7 \\
          & \multirow{2}[0]{*}{Liberia} & \multicolumn{1}{c}{NL} & 0 & 0 & 0/0 \\
          &       & \multicolumn{1}{c}{SL} & 0 & 0 & 0/0 \\
    \textbf{Total} &       &       & \textbf{9.82}  & \textbf{24.00}  & \textbf{4/11} \\
    \hline
    \multirow{6}[1]{*}{Low-High} & \multirow{3}[1]{*}{Guinea} & \multicolumn{1}{c}{UG} & 0.60 & 1.44 & 2/0 \\
          &       & \multicolumn{1}{c}{MG} & 0.60 & 1.18 & 2/0 \\
          &       & \multicolumn{1}{c}{LG} & 0.60 & 1.34 & 2/0 \\
          & Sierra Leone & \multicolumn{1}{c}{S} & 4.31 & 12.02  & 0/5 \\
          & \multirow{2}[0]{*}{Liberia} & \multicolumn{1}{c}{NL} & 1.68 & 4.91 & 2/2 \\
          &       & \multicolumn{1}{c}{SL} & 1.08 & 3.11 & 0/2 \\
    \textbf{Total} &       &       & \textbf{8.86}  & \textbf{24.00}  & \textbf{8/9} \\
    \hline
    \multirow{6}[1]{*}{High-Low} & \multirow{3}[1]{*}{Guinea} & \multicolumn{1}{c}{UG} & 1.08 & 2.44 & 0/2 \\
          &       & \multicolumn{1}{c}{MG} & 0.60 & 1.29 & 2/0 \\
          &       & \multicolumn{1}{c}{LG} & 1.08 & 2.23 & 0/2 \\
          & Sierra Leone & \multicolumn{1}{c}{S} & 6.46 & 15.47  & 0/7 \\
          & \multirow{2}[0]{*}{Liberia} & \multicolumn{1}{c}{NL} & 1.08 & 2.56 & 0/2 \\
          &       & \multicolumn{1}{c}{SL} & 0 & 0 & 0/0 \\
    \textbf{Total} &       &       & \textbf{10.29}  & \textbf{24.00}  & \textbf{2/13} \\
   \hline
    \end{tabular}%
  \label{ScenarioResult1}%
\end{table}%

The results above indicate that if the budget is tight or the disease moves fast, some countries or regions may not get the ETC allocation or treatment. For example, under the ``All High" scenario, no budget is allocated to Liberia. Therefore, in the next subsection, we introduce the equity constraint to remedy the problem of not allocating any ETCs or treatment resources to a single country or some of the regions of a country.

\subsection{Impacts of Equity Considerations}

In this subsection, we present results by adding each of the three equity constraints \eqref{cons3-ex:4}, \eqref{cons3-ex:5}, and \eqref{Prevalence Equity Constraint}, as introduced in Section \ref{Equity Constraints}, separately into the linearized multi-stage stochastic programming epidemic-logistics model \eqref{Ebola:1}. Equity constraints impose a bound on the total number of infections in each region and thus enforcing that each region considered in West Africa receives a more equitable share of resources, including ETCs and treatment funds, while minimizing the total number of infections and deaths.

According to the results, imposing the infection equity constraint \eqref{cons3-ex:4} or the prevalence equity constraint \eqref{Prevalence Equity Constraint} does not significantly change the optimal budget allocation and the total number of new infections and funerals (see Appendix \ref{Analysis of Infection Equity Constraint} for detailed results). Without introducing the infection equity constraint into the  mathematical model \eqref{Ebola:1}, the absolute value of the difference between the infection ratio and the population ratio in Guinea, Sierra Leone, and Liberia is 0.42, 0.04, and 0.38, respectively, based on the optimal solution value similar to the $k$ values considered here. This result implies that our model balances the total number of infections in each region with its population and population, even without the infection equity constraint.

\begin{figure}[H]
\centering
\includegraphics[width=6in]{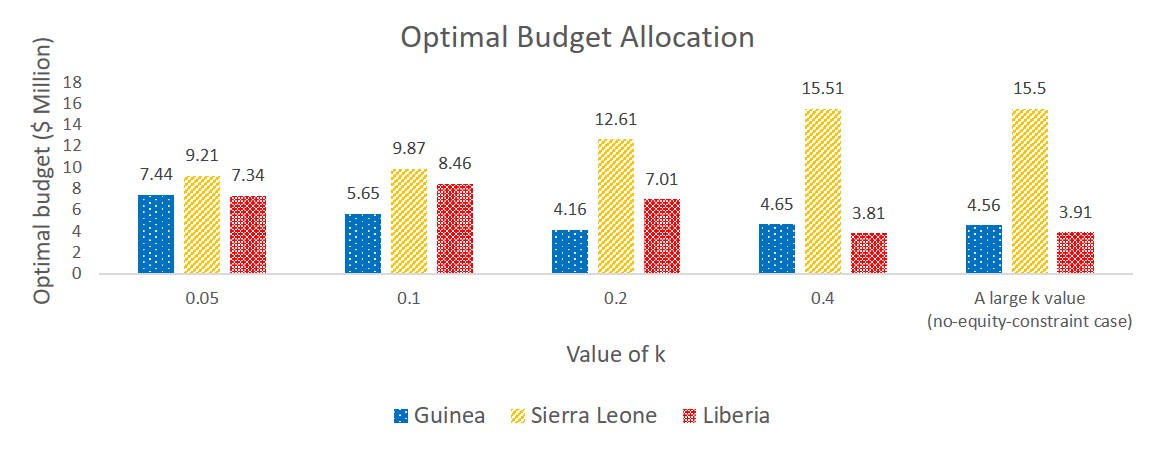}
\caption{Optimal budget allocation under different $k$ values for an 8-stage problem with \$24M budget}
\label{Fig53}
\end{figure}

\begin{figure}[H]
\centering
\includegraphics[width=6in]{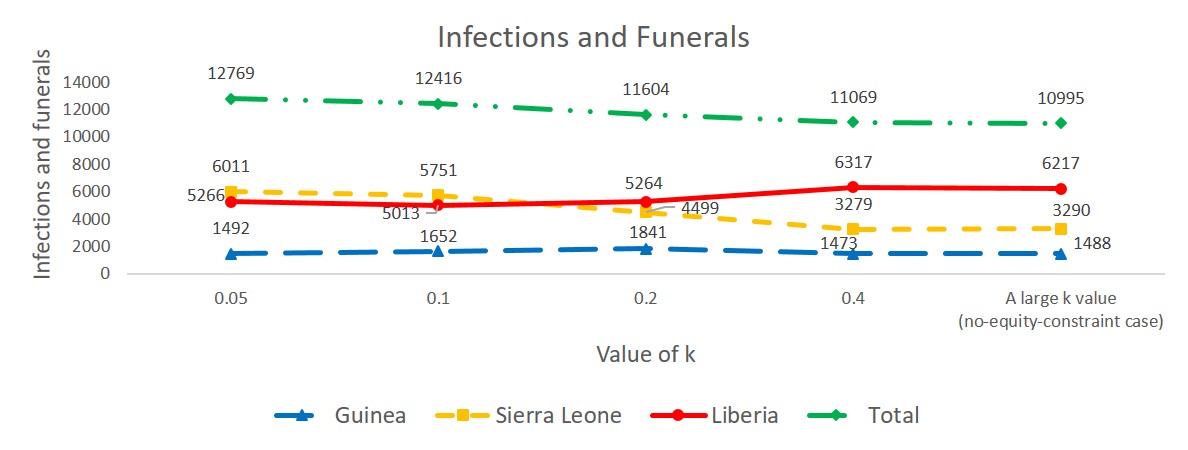}
\caption{Total number of new infections and funerals under different $k$ values for an 8-stage problem with \$24M budget}
\label{Fig54}
\end{figure}

Similar to the infection equity case, we introduce the capacity equity constraint \eqref{cons3-ex:5} into the multi-stage stochastic programming epidemic-logistics model \eqref{Ebola:1} for an 8-stage instance with the \$24M budget level under different values of $k$. Table \ref{capacity 30 gap} represents the run time specifics regarding the mathematical model \eqref{Ebola:1} with the capacity equity constraint \eqref{cons3-ex:5}, while Figures \ref{Fig53} and \ref{Fig54} present the budget allocation and the total number of infections and funerals over the three considered countries for varying $k$ values. When $k$ is larger than 0.4, we observe no significant change in the results. However, a small $k$ value can impact the results significantly. For example, when $k=0.05$, all three regions have a similar budget allocation. If $k$ increases from 0.05 to 0.2, the total number of infections and funerals in Guinea is slightly increased, but it is decreased when $k$ is further increased. Thus, allocating the majority of resources to Guinea may not be necessary, and some of those resources would be wasted. As we relax the equity capacity constraint by increasing the $k$ value from 0.05 to 0.4 and above, we observe a significant drop in the number of infected individuals and funerals in Sierra Leone. The total number of infected people and funerals over all three countries is the largest (12,769) when the capacity equity constraint is strictly enforced, and it is the smallest (10,995) when the capacity equity constraint is relaxed. This result implies that enforcing a tight equity constraint might adversely impact the total number of infections and deaths, and thus resulting in a high cost that we have to pay for fairness.

\begin{table}[H]
  \centering
	  \caption{Model run specifics with the capacity equity constraint \eqref{cons3-ex:5}}
    \begin{tabular}{ccccccccc}
    \hline
    $k$ value & Solution Time (CPU sec) & Optimality Gap (\%) \\
    \hline
     0.05 & 72,103 & 7  \\
     0.1 & 72,121 & 8  \\
     0.2 & 72,053 & 6  \\
     0.4 & 72,031 & 2  \\
     \multicolumn{1}{c}{A large $k$ value} & \multirow{2}[0]{*}{7,232} & \multirow{2}[0]{*}{0}  \\
    \multicolumn{1}{c}{(no-equity-constraint case)} & & \\
    \hline
    \end{tabular}%
  \label{capacity 30 gap}%
\end{table}%

\section{Discussion and Future Research Directions} \label{discussion and future research directions}

In this paper, we extended the epidemic-logistics model of \cite{buyuktahtakin2018new} to study an epidemic control problem in a large-scale population where the transmission rate of the disease is uncertain. To our knowledge, this is the first multi-stage stochastic epidemic-logistic model that takes into account both the uncertain disease growth and equitable resource allocation simultaneously. We consider various disease progression scenarios resulted from the realization of the community transmission rates. Our objective is to minimize the total expected number of infected individuals and funerals over all scenarios, all periods, and all regions considered. We study the value of the stochastic solution and introduce the equity constraints to analyze the fair resource allocation among different countries and multiple regions of a country. Our multi-stage VSS analysis suggests that the stochastic model considerably improves the solution of the deterministic model, and the consideration of uncertainty in a multi-stage disease-transmission model is necessary.

We define the infection level as the difference between the ratio of the number of infected people in a region to the total number of infected people over all regions and the ratio of the population in a region to the total population over all regions. Under tight budget levels, most of the budget would be allocated to the region that has the highest initial infection level, while other regions would receive ETCs and treatment resources according to their infection level as the available budget increases. This indicates that the initial infection level is a key factor in resource allocation. Additionally, more 100-bed ETCs would be allocated to the country that has a high infection level since more capacity will be needed to treat infected people while saving from the fixed cost of opening new ETCs.

According to the results, our model allocated most of ETCs in the first stage to provide a quick response to the epidemic and reduce a large number of unnecessary infections and funerals. Our results showed that the number of untreated infections dropped quickly when early actions were taken with a sufficiently large budget, and the disease was controlled much faster than the report date of the World Health Organization (WHO). The uncertainty in disease transmission is a critical factor that makes it challenging to manage an outbreak in a real-life situation. To be more specific, the transmission rate might suddenly become high after a latent period, and the existing resources may not be sufficient to handle such unexpected situations. Consequently, a large number of unisolated and untreated individuals could stay in the community and continue to spread the disease, as in the case of the current outbreak of Coronavirus (COVID-19) disease \citep{WHOCoronavirus}. Thus, the preparedness and early action to handle the uncertain disease transmission are crucial, and we would rather ``the beds waiting for people'' than ``people waiting for the beds.'' Our findings are consistent with several other articles that also report the importance of early action for epidemic control \citep{lekone2006statistical, jacobsen2016lessons, siedner2015strengthening}. The lessons learned from the EVD control in West Africa by WHO and Centers for Disease Control and Prevention (CDC) also indicate that an early action will have a significant improvement in slowing down an epidemic and eventually stopping it \citep{Esummary, WHO}.

Different than the former literature, the solutions of our multi-stage stochastic programming model show that the optimal timing of the resource allocation might vary if we have a relatively ample budget. For instance, in both \$24M and \$48M budget levels, some resources were allocated throughout the planning horizon in some locations, such as Guinea and Sierra Leone. This is because we have more budget to take action when the transmission of the disease gets worse. This result shows that the timing of the resource allocation should be decided dynamically and based on the predicted disease growth scenario and budget, and thus implying the superiority of a multi-stage stochastic programming model over a two-stage or static model again.

We analyzed five specific disease growth scenarios and studied resource allocation strategies under each scenario. Under the scenarios in which the disease moves faster, more number of ETCs are allocated compared to the scenarios in which the disease moves slower to treat more people. In addition, if the disease moves faster, the majority of the capacity is allocated to the region that has the highest initial infection level.  If the disease consistently moves at a slow rate, the treatment capacity is allocated more equally among regions to help fight against the disease.  In the ``Low-High'' case, in which the disease moves in a slow rate first and then starts to be more aggressive in the following time stages, the model allocates budget immediately to the regions with a high infection level and knocks down the number of infected individuals to low values, which will lessen the impacts of a high disease transmission rate later in the planning horizon. Because an initially slow-moving disease gives us more time to get prepared to control the disease spread, the ``Low High'' case can be considered as a better scenario compared to the ``High-Low'' case.

We introduced the infection and capacity equity constraints separately into our model to analyze the impact of enforcing fairness in resource allocation. Solutions obtained with the infection equity constraint imply that the original optimal solution balances the resource allocation among multiple regions in a similar fashion to the infection equity constraint. Thus, our model takes into account the ratio of infection to the total infection level as well as the ratio of the population to the total population level over all three countries while making the resource allocation decision.

When a tight capacity equity constraint is enforced, the budget is allocated equally to the three regions. However, in this case, some of the budget may be wasted, and no obvious effects are brought out by providing additional capacity to a region based solely on its population.  This result shows that allocating treatment resources proportional to population is sub-optimal, which is also consistent with the findings of  \cite{ren2013optimal}. When the capacity equity constraint is relaxed, the number of infections and funerals in Guinea and Liberia is slightly changed, but this number decreased significantly for Sierra Leone, and the total three countries. For both tight and ample budget cases, the total number of infections and funerals is much higher when the capacity equity constraint is strictly forced, resulting in a heavy price we would have to pay for perfect equity in resource allocation. This result implies that the decision maker should be cautious about enforcing fairness when allocating resources to multiple regions.

There are several important future research directions that arise out of this study. For example, the impact of vaccinations currently used to prevent the spread of the disease could be analyzed in a future study. The influence of vaccination is group-specific, and thus susceptible individuals can be divided into different groups according to their age, sex, race, and health status. Due to the lack of available data, the transmission rate from susceptible individuals to infected individuals would be more difficult to predict under vaccination. Furthermore, different kinds of vaccines used, the amount of vaccine allocated to each region, and the time when vaccination becomes accessible might impact the disease transmission rate significantly. Our model could be extended by adding a compartmental class named as ``vaccinated'' to study the various dimensions of vaccination.

Moreover, our multi-stage stochastic program only includes the expectation criterion in the objective function when it compares random variables to find the best decisions. Thus, our study provides a risk-neutral approach. In a future extension of this work, risk measures, such as Conditional Value at Risk ($CVaR$), could be incorporated into the objective function to reflect the perspectives of a risk-averse decision maker.

\section*{Acknowledgments} 
The authors thank two anonymous referees, the associate editor, and the editor, whose remarks helped to improve the content and clarity of our exposition.

\begin{appendices}

\section{Notation}
\label{Notation}

Model notations that are used throughout the rest of this paper are presented in Tables~\ref{Sets}--\ref{Decision} below.

\begin{table}[H]
\small
  \centering
  \caption{Sets and indices.}
    \begin{tabular}{cllllllll}
    \hline
    Notation & Description \\
    \hline
      $J$ & Set of time periods, $J=\lbrace0,...,\overline{J}\rbrace$. \\
$A$ & Set of ETC types, $A=\lbrace1,...,\overline{A}\rbrace$.\\
$R$ & Set of regions, $R=\lbrace1,...,\overline{R}\rbrace$.\\
$M_{r}$ & Set of all surrounding regions of region $r$.\\
$\Omega$ & Set of scenarios, $\Omega=\lbrace1,...,\overline{\Omega}\rbrace$.\\
$j$ & Index for time period where $j \in J$.\\
$r$ & Index for region where $r\in R$.\\
$a$ & Index defining type of ETC, where $a\in A$.\\
$\omega$ & Index for scenario where $\omega \in \Omega$.\\
    \hline
    \end{tabular}%
  \label{Sets}%
\end{table}%

\begin{table}[H]
\small
  \centering
  \caption{Transition parameters describing the rate of movement between disease compartments.}
    \begin{tabular}{cllllllll}
    \hline
    Notation & Description \\
    \hline
      $\lambda_{1,r}$ & Disease fatality rate without treatment in region $r$.\\
$\lambda_{2,r}$ & Disease fatality rate while receiving treatment in region $r$.\\
$\lambda_{3,r}$ & Disease survival rate without treatment in region $r$.\\
$\lambda_{4,r}$ & Disease survival rate with treatment in region $r$.\\
$\lambda_{5,r}$ & Safe burial rate of Ebola-related dead bodies in region $r$.\\
$\chi_{1,r}^{\omega}$ & Transmission rate per person due to community interaction in region $r$\\
& under scenario $\omega$.\\
$\chi_{2,r}$ & Transition rate per person during traditional funeral ceremony in region $r$.\\
    \hline
    \end{tabular}%
  \label{Transit}%
\end{table}%

\begin{table}[H]
\small
  \centering
  \caption{Other parameters.}
    \begin{tabular}{cllllllll}
    \hline
    Notation & Description \\
    \hline
      $b_{1j,r}$ & Unit cost of treatment for an infected individual in region $r$ at end of period $j$.\\
$b_{2j,r}$ & Unit cost of safe burial for a dead body in region $r$ at end of period $j$.\\
$g_{aj,r}$ & Fixed cost of establishing type $a$ ETC in region $r$ at end of period $j$.\\
$k_a$ & Capacity (number of beds) of type $a$ ETC.\\
$u_r$ & The population in region $r$.\\
$\Delta$ & Total available budget for treatment. \\
$\pi_{r}$ & Initial number of susceptible individuals in region $r$.\\
$\varpi_{r}$ & Initial number of infected individuals in region $r$.\\
$\theta_{r}$ & Initial number of treated individuals in region $r$.\\
$\vartheta_{r}$ & Initial number of recovered individuals in region $r$.\\
$\upsilon_{r}$ & Initial number of unburied dead bodies (funerals) in region $r$.\\
$\tau_{r}$ & Initial number of buried dead bodies (safe burials) in region $r$.\\
$\varsigma_{r}$ & Initial treatment capacity in terms of number of ETC beds in region $r$.\\
$\varepsilon_{l \rightarrow r}$ & Migration rate of susceptible individuals from surrounding regions $l \in M_{r}$ to region $r$.\\
$\phi_{l \rightarrow r}$ & Migration rate of infected individuals from surrounding regions $l \in M_{r}$  to region $r$.\\
$\nu_{r \rightarrow l}$  & Migration rate of susceptible individuals from region $r$ to surrounding regions $l \in M_{r}$.\\
$\rho_{r \rightarrow l}$ & Migration rate of infected individuals from region $r$ to surrounding regions $l \in M_{r}$.\\
    \hline
    \end{tabular}%
  \label{Other}%
\end{table}%

\begin{table}[H]
\small
  \centering
  \caption{State variables.}
    \begin{tabular}{cllllllll}
    \hline
    Notation & Description \\
    \hline
      $S_{j,r}^{\omega}$ & Number of susceptible individuals in region $r$ at end of period $j$ under scenario $\omega$.\\
$I_{j,r}^{\omega}$ & Number of infected individuals in region $r$ at end of period $j$ under scenario $\omega$.\\
$T_{j,r}^{\omega}$ & Number of individuals receiving treatment in region $r$ at end of period $j$\\
& under scenario $\omega$.\\
$R_{j,r}^{\omega}$ & Number of recovered individuals in region $r$ at end of period j under scenario $\omega$.\\
$F_{j,r}^{\omega}$ & Number of deceased individuals due to the epidemic in region $r$ at end of period $j$\\
& under scenario $\omega$.\\
$B_{j,r}^{\omega}$ & Number of buried individuals in region $r$ at end of period $j$ under scenario $\omega$.\\
$\hat{S}_{j,r}^{\omega}$ & Number of susceptible individuals migrating into region $r$ at end of period $j$\\
& under scenario $\omega$.\\
$\widetilde{S}_{j,r}^{\omega}$ & Number of susceptible individuals emigrating from region $r$ at end of period $j$\\
& under scenario $\omega$.\\
$\hat{I}_{j,r}^{\omega}$ & Number of infected individuals migrating into region $r$ at end of period $j$\\
& under scenario $\omega$.\\
$\widetilde{I}_{j,r}^{\omega}$ & Number of infected individuals emigrating from region $r$ at end of period $j$\\
& under scenario $\omega$.\\
    \hline
    \end{tabular}%
  \label{State}%
\end{table}%

\begin{table}[H]
\small
  \centering
  \caption{Decision variables.}
    \begin{tabular}{cllllllll}
    \hline
    Notation & Description \\
    \hline
      $C_{j,r}^{\omega}$ & Total capacity (number of beds) of established ETCs in region $r$ at end of period $j$\\
& under scenario $\omega$.\\
$\overline{I}_{j,r}^{\omega}$ & Number of infected individuals hospitalized (and quarantined) in region $r$\\
& at end of period $j$ under scenario $\omega$.\\
$y_{aj,r}^{\omega}$ & Number of type $a$ ETCs established in region $r$ at end of period $j$\\
& under scenario $\omega$.\\
    \hline
    \end{tabular}%
  \label{Decision}%
\end{table}%

\section{Non-Anticipativity}
\label{Non-Anticipativity}

Two scenarios should have the same decision variables at a stage $j$ if they share the same scenario path up to that stage. Corresponding decisions up to stage $j$ of two inseparable scenarios should be the same. These implications are named as non-anticipativity constraints, and can be formulated as follows. Consider the node marked $n$ in the scenario tree, and denote the corresponding stage as $t(n)$. Let the set of scenarios that pass through node $n$ be $\beta(n)$. We must ensure that decision variables at stage $t(n)$ that are associated with node $n$ (for example: $x_{t(n)}^{\omega}$) have identical values for $\omega \in \beta(n)$. One way to do this is to add the non-anticipativity constraint as in the following form:

\begin{equation}
x_{t(n)}^{\omega}-x_n=0 \qquad \forall \omega \in \beta(n). \nonumber
\end{equation}

As an example, consider the first three stages of the multi-stage problem shown in Figure \ref{Scheme}. The set of nodes of this scenario tree is given by $N=\lbrace 0, 1, 2, 3, ..., 13, 14 \rbrace$, where $t(0)=0, t(1)=1, t(2)=1, t(3)=2, t(4)=2, t(5)=2, t(6)=2, t(7)=t(8)=t(9)=t(10)=t(11)=t(12)=t(13)=3$. The set of scenarios that share node $n=2$ is given by $\beta(2)=\lbrace 5, 6, 7, 8\rbrace$. Let $x_{t(2)}^{\omega}$ represent decision variables for $\omega \in \beta(2)$. The non-anticipativity constraint for those variables can be written as:
\begin{equation}
x_{t(2)}^{\omega}-x_2=0 \qquad \forall \omega \in \beta(2). \nonumber
\end{equation}

\section{Linearization}
\label{Linearization}
%

We first linearize the logical constraint that describes the number of hospitalized individuals in equation \eqref{cons3-ex:3}. Following the method of \cite{kibics2017optimizing}, for each $j\in J\setminus \overline{J}$, $r\in R$, and $\omega\in \Omega$, constraint \eqref{cons3-ex:3} can be written as:
\begin{equation} 
\overline{I}_{j,r}^{\omega}=(C_{j,r}^{\omega}-T_{j,r}^{\omega})z_{j,r}^{\omega}+I_{j,r}^{\omega}(1-z_{j,r}^{\omega}), \label{lin1-ex:1}
\end{equation} 
where $z_{j,r}^{\omega}$ is a binary variable, which takes the value $1$ if the number of infected individuals to be hospitalized is restricted by the number of available beds in ETCs, and the value $0$ if the number of beds in ETCs is sufficiently large to hospitalize all infected individuals. In order to ensure that $\overline{I}_{j,r}^{\omega}$ takes the minimum value of $(C_{j,r}^{\omega}-T_{j,r}^{\omega})$ and $I_{j,r}^{\omega}$, we should have the following inequalities satisfied for each $j\in J\setminus \overline{J}$, $r\in R$, and $\omega\in \Omega$:
\begin{subequations}
\label{Ebola:2}
\begin{eqnarray}
   &&  \overline{I}_{j,r}^{\omega} \leq C_{j,r}^{\omega}-T_{j,r}^{\omega} \\
   &&  \overline{I}_{j,r}^{\omega}\leq I_{j,r}^{\omega}.
\end{eqnarray}
\end{subequations}
However, constraint \eqref{lin1-ex:1} is still non-linear due to quadratic terms. Therefore, two auxiliary variables $U_{j,r}^{\omega}$ and $W_{j,r}^{\omega}$ are introduced to be substituted with $(C_{j,r}^{\omega}-T_{j,r}^{\omega})z_{j,r}^{\omega}$ and $I_{j,r}^{\omega}(1-z_{j,r}^{\omega})$, respectively. In this case, for each $j\in J\setminus \overline{J}$, $r\in R$, and $\omega\in \Omega$, constraint \eqref{lin1-ex:1} can be written as:
\begin{equation} 
\overline{I}_{j,r}^{\omega}=U_{j,r}^{\omega}+W_{j,r}^{\omega} \label{lin1-ex2}
\end{equation} 

We then introduce a lower bound ($H_{LB}$) and upper bound ($H_{UB}$) for  $C_{j,r}^{\omega}-T_{j,r}^{\omega}$, such that $H_{LB}\leq C_{j,r}^{\omega}-T_{j}^{\omega} \leq H_{UB}$ and add the following constraints to the model for each $j\in J\setminus \overline{J}$, $r\in R$, and $\omega\in \Omega$:
\begin{subequations}
\label{Ebola:3}
\begin{eqnarray} 
   &&  U_{j,r}^{\omega}\leq H_{UB}z_{j,r}^{\omega} \label{lin1-ex3}\\
   &&  U_{j,r}^{\omega}\geq H_{LB}z_{j,r}^{\omega}\\
   &&  U_{j,r}^{\omega}\leq (C_{j,r}^{\omega}-T_{j,r}^{\omega})-H_{LB}(1-z_{j,r}^{\omega})\\
   &&  U_{j,r}^{\omega}\geq (C_{j,r}^{\omega}-T_{j,r}^{\omega})-H_{UB}(1-z_{j,r}^{\omega}) \label{lin1-ex4}.
\end{eqnarray}
\end{subequations} 

Similarly, we introduce a lower bound ($I_{LB}$) and an upper bound ($I_{UB}$) for $I_{j,r}^{\omega}$, such that $I_{LB}\leq I_{j,r}^{\omega} \leq I_{UB}$, and add the following four constraints for each $j\in J\setminus \overline{J}$, $r\in R$, and $\omega\in \Omega$ to the model:
\begin{subequations}
\label{Ebola:3}
\begin{eqnarray} 
   &&  W_{j,r}^{\omega}\leq I_{UB}(1-z_{j,r}^{\omega}) \label{lin1-ex5} \\
   &&  W_{j,r}^{\omega}\geq I_{LB}(1-z_{j,r}^{\omega})\\
   &&  W_{j,r}^{\omega}\leq I_{j,r}^{\omega}-I_{LB}z_{j,r}^{\omega}\\
   &&  W_{j,r}^{\omega}\geq I_{j,r}^{\omega}-I_{UB}z_{j,r}^{\omega} \label{lin1-ex6}.
\end{eqnarray}
\end{subequations}

Thus the constraint \eqref{cons3-ex:3} can be equivalently linearized by replacing it with constraints \eqref{lin1-ex2}, \eqref{lin1-ex3}-\eqref{lin1-ex4} and \eqref{lin1-ex5}-\eqref{lin1-ex6}.

We then linearize the equity constraint given by equation \eqref{cons3-ex:4}. By multiplying the two denominators on the left side of \eqref{cons3-ex:4} by each other and multiplying the right side of \eqref{cons3-ex:4} by $\sum\limits_{r \in R}u_r\sum\limits_{j \in J}\sum\limits_{r \in R}\sum\limits_{\omega \in W}P^\omega I_{j,r}^{\omega}$, we obtain the following inequality:
\begin{equation} 
\vert \sum\limits_{r \in R}u_r\sum\limits_{j \in J}\sum\limits_{\omega \in W}P^\omega I_{j,r}^{\omega} - u_{r}\sum\limits_{j \in J}\sum\limits_{r \in R}\sum\limits_{\omega \in W}P^\omega I_{j,r}^{\omega} \vert \leq k \sum\limits_{r \in R}u_r\sum\limits_{j \in J}\sum\limits_{r \in R}\sum\limits_{\omega \in W}P^\omega I_{j,r}^{\omega}
\label{lin2-ex:1}
\end{equation} 
The absolute value in inequality \eqref{lin2-ex:1} could be linearized using the following two constraints:
\begin{subequations}
\label{Ebola:4}
\begin{eqnarray} 
   &&  \sum\limits_{r \in R}u_r\sum\limits_{j \in J}\sum\limits_{\omega \in W}P^\omega I_{j,r}^{\omega} - u_{r}\sum\limits_{j \in J}\sum\limits_{r \in R}\sum\limits_{\omega \in W}P^\omega I_{j,r}^{\omega} - k\sum\limits_{r \in R}u_r\sum\limits_{j \in J}\sum\limits_{r \in R}\sum\limits_{\omega \in W}P^\omega I_{j,r}^{\omega} \leq 0 \quad \quad \label{lin2-ex2} \\
   &&  \sum\limits_{r \in R}u_r\sum\limits_{j \in J}\sum\limits_{\omega \in W}P^\omega I_{j,r}^{\omega} - u_{r}\sum\limits_{j \in J}\sum\limits_{r \in R}\sum\limits_{\omega \in W}P^\omega I_{j,r}^{\omega} + k\sum\limits_{r \in R}u_r\sum\limits_{j \in J}\sum\limits_{r \in R}\sum\limits_{\omega \in W}P^\omega I_{j,r}^{\omega} \geq 0 \quad \quad \label{lin2-ex3}
\end{eqnarray}
\end{subequations}
Therefore, the constraint \eqref{cons3-ex:4} can be equivalently linearized by replacing it with constraints \eqref{lin2-ex2} and \eqref{lin2-ex3}. 

Similarly, we have replaced the non-linear capacity equity constraint \eqref{cons3-ex:5} with the following two linear constraints:
\begin{subequations}
\label{Ebola:5}
\begin{eqnarray} 
   &&  \sum\limits_{r \in R}u_r\sum\limits_{j \in J}\sum\limits_{\omega \in W}P^\omega C_{j,r}^{\omega} - u_{r}\sum\limits_{j \in J}\sum\limits_{r \in R}\sum\limits_{\omega \in W}P^\omega C_{j,r}^{\omega} - k\sum\limits_{r \in R}u_r\sum\limits_{j \in J}\sum\limits_{r \in R}\sum\limits_{\omega \in W}P^\omega C_{j,r}^{\omega} \leq 0 \quad \quad \label{lin2-ex4} \\
   &&  \sum\limits_{r \in R}u_r\sum\limits_{j \in J}\sum\limits_{\omega \in W}P^\omega C_{j,r}^{\omega} - u_{r}\sum\limits_{j \in J}\sum\limits_{r \in R}\sum\limits_{\omega \in W}P^\omega C_{j,r}^{\omega} + k\sum\limits_{r \in R}u_r\sum\limits_{j \in J}\sum\limits_{r \in R}\sum\limits_{\omega \in W}P^\omega C_{j,r}^{\omega} \geq 0 \quad \quad \label{lin2-ex5}
\end{eqnarray}
\end{subequations}

The non-linear multi-stage stochastic programming epidemic–logistics model \eqref{Ebola:1} is converted into an equivalent MIP formulation by replacing the non-linear capacity availability constraint \eqref{cons3-ex:3} with constraints \eqref{lin1-ex2}, \eqref{lin1-ex3}-\eqref{lin1-ex4} and \eqref{lin1-ex5}-\eqref{lin1-ex6}, the non-linear infection equity constraint \eqref{cons3-ex:4} with constraints \eqref{lin2-ex2} and \eqref{lin2-ex3}, and the non-linear capacity equity constraint \eqref{cons3-ex:5} with constraints \eqref{lin2-ex4} and \eqref{lin2-ex5}. In the next section, we present a case study involving the control of the 2014–2015 Ebola outbreak in the three most-affected West African countries, Guinea, Sierra Leone, and Liberia.

\section{Ebola Case Study Data}
\label{Ebola Case Study Data}

This section presents the data used to formulate the model, including population and migration data, resource cost data, and epidemiological data. All data provided in this section was collected using literature resources and given bi-weekly. Data pertaining to the 2014-2015 Ebola outbreak and the deterministic epidemics-logistics model have been validated by \cite{buyuktahtakin2018new}. 

\subsection{Population and Migration Data}
\label{Population and Migration Data}
Table \ref{populationdata} presents the distribution of the population in Guinea, Liberia, and Sierra Leone, all located in West Africa. We consider six regions: three of them are located in Guinea (Upper Guinea (UG), Middle Guinea (MG), and Lower Guinea (LG)), two of them are in Liberia (Northern Liberia (NL) and Southern Liberia (SL)) and the last one, Sierra Leone, is a county itself (S). Table \ref{initial infection} shows the total number of initial infections in each country. Table \ref{migrationdata} gives the migration rates from each of the five regions (UG,MG,LG,NL,SL) to the other four regions. There is no migration in Sierra Leone because it is considered as a region by itself. Rapidly after the initial recognition of the Ebola outbreak, those three countries closed the national borders, so we only consider the migration within a country.
\begin{table}[H]\footnotesize
  \centering
	\renewcommand\arraystretch{0.9}
\renewcommand\tabcolsep{3.0pt}
  \caption{Regions, population size and rate in West Africa  }
    \begin{tabular}{lllllllll}
    \hline
    Guinea & Population & Ratio & Liberia & Population & Ratio & Sierra Leone & Population & Ratio \\
					 & (millions) &       &         &  (millions) &      &              &  (millions) &  \\
    \hline
    UG     & 4,3 & 0.41  & NL     & 2,2  & 0.64  & S & 4,9 & 1.00 \\
    MG     & 2,7 & 0.25  & SL     & 1,2  & 0.36  &       &       &  \\
    LG     & 3,7 & 0.34  &       &       &       &       &       &  \\
    \textbf{Total} & \textbf{10,7 } & 1.00  &       & \textbf{3,4} & 1.00  &       & \textbf{4,9 } & 1.00 \\
    \hline
    \end{tabular}%
  \label{populationdata}%
\end{table}%

\begin{table}[H]
  \centering
  \caption{The number of infected people at the beginning of the planning horizon (August 30, 2014) in West Africa}
    \begin{tabular}{ccccccccc}
    \hline
    Guinea & Sierra Leone & Liberia \\
    \hline
     218 & 604 & 685  \\
    \hline
    \end{tabular}%
  \label{initial infection}%
\end{table}%

\begin{table}[H]\footnotesize
  \centering
  \caption{Bi-weekly migration rate between regions of Guinea and Liberia, original data acquired from \cite{wesolowski2014commentary}.}
    \begin{tabular}{lllllllll}
    \hline
 From $\setminus$ To    & UG & MG & LG & NL & SL\\
    \hline
    UG     & & 0.0032 & 0.0010 \\
    MG     & 0.0052 & & 0.0025\\
    LG     & 0.0012 & 0.0018\\
    NL & & & & & 0.0007\\
    SL & & & & 0.0011\\
    \hline
    \end{tabular}%
  \label{migrationdata}%
\end{table}%

\subsection{Resource Allocation Cost Data}
\label{Resource Allocation Cost Data}
The fixed cost of locating Ebola treatment centers (ETCs) and the per-person cost of Ebola treatment for either 50 or 100-bed ETC are given below in Table \ref{costdata}. The treatment cost includes the fixed cost for establishing each type of ETCs, isolation unit center, and laboratory diagnosis. Additionally, each facility has a variable running cost mainly composed of treating infected people and contact tracing of the infected individuals. There is also a safe burial cost for safely burying infected dead bodies. Fixed costs are one-time; however, all other costs are given for a 2-week period in Table \ref{costdata}. For example, the variable cost of the Ebola treatment center represents the cost of treating one infected individual over two weeks.

\begin{table}[H]\footnotesize
\begin{threeparttable}
  \centering
  \caption{Summary of Ebola treatment cost for 50 (100)-bed ETC.}
    \begin{tabular}{lllll}
    \hline
    Cost description & Fixed Cost &   & Variable cost\tnote{*}  & Safe burial cost\tnote{*}  \\
    \hline
    Ebola treatment center& \$386,000 (\$694,800)& & \$8,810 &  \\
    Isolation unit center (IUC) & \$112,500  &   & \$1,133 &  \\
    Laboratory diagnosis & \$100,000  &   & \$540  &  \\
    Subnational technical services &   &   & \$2,250  &  \\
    Contact tracing &   &   & \$1,128  &  \\
    Safe burial &   &   &   & \$1,127  \\
    \textbf{Total} & \textbf{\$598,500} (\textbf{\$1,077,300})& &  \textbf{\$13,860} & \textbf{\$1,127} \\
    \hline
    \end{tabular}%
  \label{costdata}%
	\begin{tablenotes}[para]
		\footnotesize
			\item[*] {Variable and safe burial costs are bi-weekly}.								
\end{tablenotes}
\end{threeparttable}
\end{table}%

\subsection{Epidemiological Data}
\label{Epidemiological data}

Table \ref{epidemicdata} presents the data values for transmission parameters for each of the three considered countries in West Africa. The data contains the fatality rate with and without treatment, recovery rate with and without treatment, safe burial rate, and transmission rates. Because the transmission rate in the community is an uncertain parameter, we present its value under each of the two realizations as low and high. Moreover, we considered the expected value of the transmission rate at a traditional funeral for each country.

\begin{table}[H]\footnotesize
  \centering
		\renewcommand\arraystretch{0.9}
\renewcommand\tabcolsep{0.5pt}
  \caption{Transmission parameters and bi-weekly rates for the Ebola outbreak.}
    \begin{tabular}{llllll}
    \hline
    Parameter & Description & \ \ \ \ \ Data  &       &       & Reference \\
  & & Guinea & Sierra Leone & Liberia & \\
    \hline
    $\lambda_1$    & Rate of fatality without treatment &    0.428   &      0.124 &   0.176    &  \cite{WHO}, \cite{who2014ebola}\\
    $\lambda_2$     & Rate of fatality with treatment &   0.350    &      0.096 &    0.128   &  \cite{WHO}, \cite{who2014ebola}\\
    $\lambda_3$     & Rate of recovery without treatment &  0.240     &      0.242 &    0.232   &  \cite{WHO}, \cite{who2014ebola}\\
    $\lambda_4$     & Rate of recovery with treatment &   0.416    &    0.327   &    0.312   &  \cite{WHO}, \cite{who2014ebola}\\
    $\lambda_5$     & Safe burial rate &   0.730    &    0.710   &    0.740   &  \cite{WHO}, \cite{who2014ebola}\\
    $\chi_{1,r}^{l}$     & Transmission rate in community (Low) &   0.660    &   0.632    &   0.560    &  \cite{camacho2014potential}\\
    $\chi_{1,r}^{h}$     & Transmission rate in community (High) &   0.990    &    0.940   &    0.840   &  \cite{camacho2014potential}\\
    $\chi_{2,r}$    & Transmission rate at traditional funeral & 1.460  & 1.420  & 1.480  &  \cite{camacho2014potential}\\
    \hline
    \end{tabular}%
  \label{epidemicdata}%
\end{table}%

\section{Analysis of Infection and Prevalence Equity Constraints}
\label{Analysis of Infection Equity Constraint}

The infection equity constraint \eqref{cons3-ex:4} limits the difference between the proportion of infections in each region over the total number of infections and the proportion of the population at each region over the total population with a specific $k$ value. Introducing the infection equity constraint to the mathematical model with 8 stages increased the average CPU solution time from 7200 seconds to 10 hours when $k=0.2$, and the average optimality gap from 1\% to 29\%. Table \ref{infection gap} gives the run time specifics regarding the mathematical model \eqref{Ebola:1} with eight stages and the infection equity constraint \eqref{cons3-ex:4}. As seen in Table \ref{infection gap}, for $k$ values between 0.2 and 0.4, the computational complexity significantly increases compared to the case where the infection equity constraint is relaxed, i.e., $k$ is set to a large number.

\begin{figure}[H]
\centering
\includegraphics[width=6in]{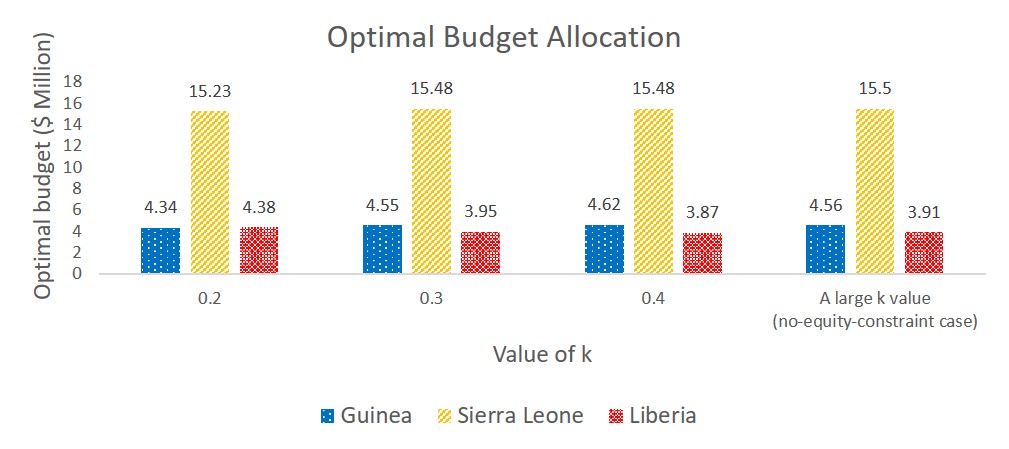}
\caption{Optimal budget allocation under different $k$ values for an 8-stage problem with \$24M budget}
\label{Fig49}
\end{figure}

\begin{figure}[H]
\centering
\includegraphics[width=6in]{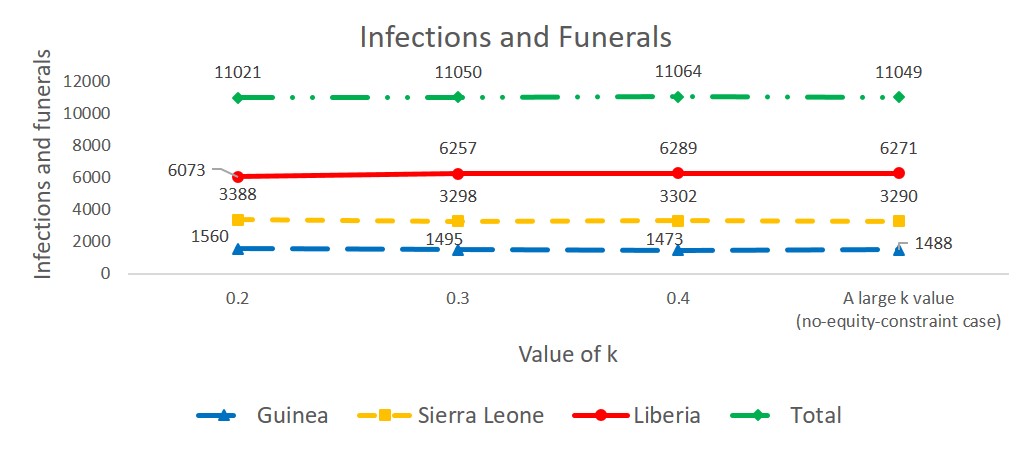}
\caption{Total number of new infections and funerals under different $k$ values for an 8-stage problem with \$24M budget} 
\label{Fig50}
\end{figure}

Figures \ref{Fig49} and \ref{Fig50} show the budget allocation and the total number of infections and funerals over the three considered countries for different $k$ values. According to the results, varying $k$ values does not significantly change the optimal budget allocation and the total number of infections and funerals. Without introducing the infection equity constraint into the  mathematical model \eqref{Ebola:1}, the absolute value of the difference between the infection ratio and the population ratio in Guinea, Sierra Leone, and Liberia is 0.42, 0.04, and 0.38, respectively, based on the optimal solution value similar to the $k$ values considered here. 


\begin{table}[H]
  \centering
  \caption{Model run specifics with the infection equity constraint \eqref{cons3-ex:4}}
    \begin{tabular}{ccccccccc}
    \hline
    $k$ value & Solution Time (CPU sec) & Optimality Gap (\%) \\
    \hline
     0.2 & 36,068 & 29  \\
     0.3 & 7,213 & 1  \\
     0.4 & 7,214 & 1  \\
    \multicolumn{1}{c}{A large $k$ value} & \multirow{2}[0]{*}{7232} & \multirow{2}[0]{*}{0}  \\
    \multicolumn{1}{c}{(no-equity-constraint case)} & & \\
    \hline
    \end{tabular}%
  \label{infection gap}%
\end{table}%

As a comparison, we also test the prevalence equity constraint \eqref{Prevalence Equity Constraint} and compare it to the infection equity constraint \eqref{cons3-ex:4}. The prevalence equity constraint bounds the absolute difference between the regional prevalence (cases per population in a region) and the country prevalence (cases per population over all regions). Without the prevalence equity constraint \eqref{Prevalence Equity Constraint} constraint, the absolute value of the difference between the infection ratio over a region and the infection ratio over all regions in Guinea, Sierra Leone, and Liberia is  $4.4\times10^{-4}$, $8.2\times10^{-5}$, and $1.2\times10^{-3}$, respectively, based on the optimal solution value.

\begin{figure}[H]
\centering
\includegraphics[width=6in]{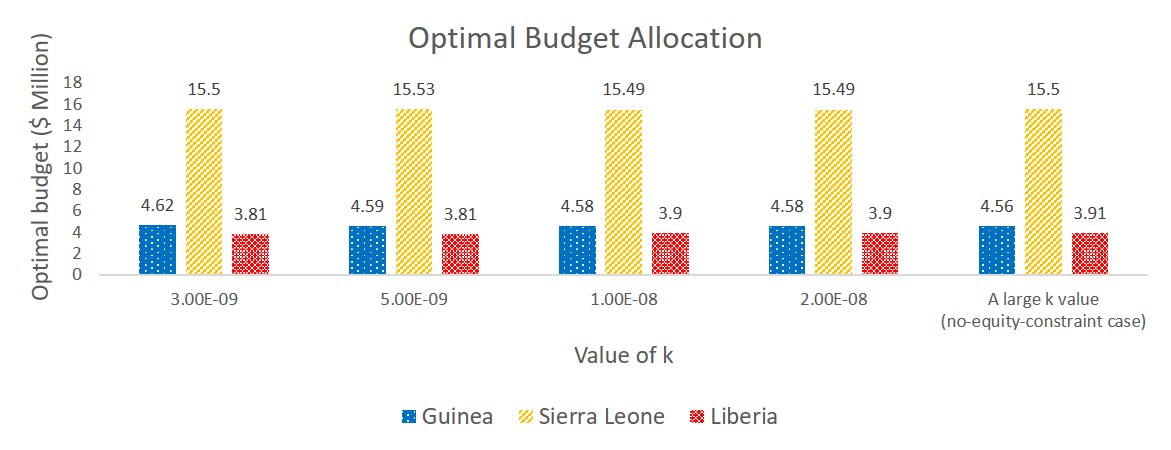}
\caption{Optimal budget allocation under different $k$ values for an 8-stage problem with \$24M budget}
\label{PrevalenceBud}
\end{figure}

\begin{figure}[H]
\centering
\includegraphics[width=6in]{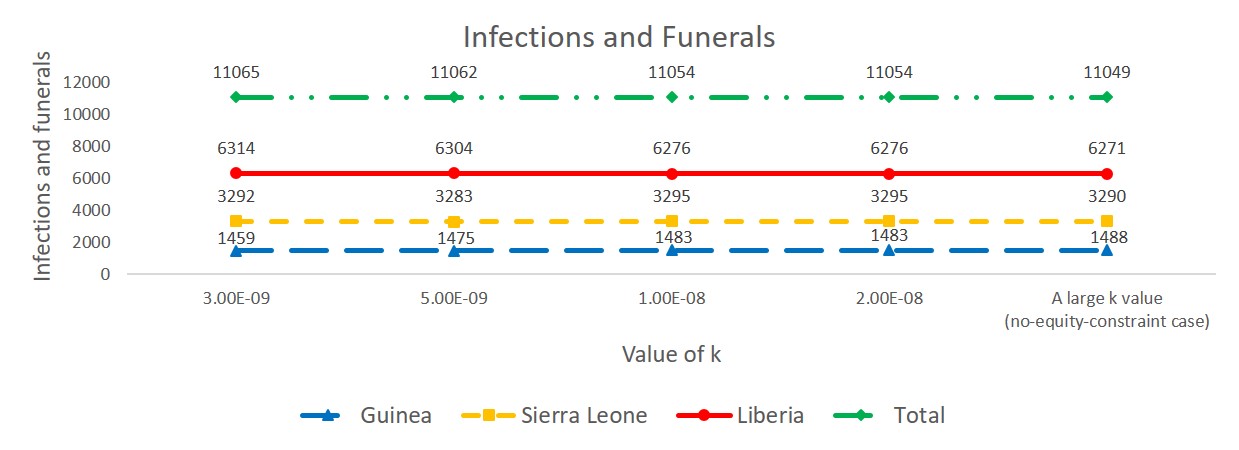}
\caption{Total number of new infections and funerals under different $k$ values for an 8-stage problem with \$24M budget}
\label{PrevalenceNum}
\end{figure}

We test the prevalence equity constraint under the \$24M budget level. Table \ref{prevalence gap} presents the run-time and optimality gap specifics for each $k$ value in inequality \eqref{Prevalence Equity Constraint} . Figures \ref{PrevalenceBud} and \ref{PrevalenceNum} show the optimal budget allocation and the number of infections and funerals under different $k$ values, respectively. Note that the $k$ values used in the prevalence equity constraint \eqref{Prevalence Equity Constraint} are much smaller than the $k$ values used in the infection equity constraint \eqref{cons3-ex:4}. Similar to the infection equity constraint, the optimal budget allocation does not show any significant difference among each $k$ value, but the number of infections and funerals slightly reduces when we relax the prevalence equity constraint. These results imply that our model balances the proportion of infections in each region, even without imposing the infection equity \eqref{cons3-ex:4} or \eqref{Prevalence Equity Constraint} prevalence equity constraints.

\begin{table}[H]
  \centering
  \caption{Model run specifics with the prevalence equity constraint}
    \begin{tabular}{ccccccccc}
    \hline
    $k$ value & Solution Time (CPU sec) & Optimality Gap (\%) \\
    \hline
     $3\times10^{-9}$ & 36,041 & 1  \\
     $5\times10^{-9}$ & 7,204 & 1  \\
     $1\times10^{-8}$ & 7,232 & 1  \\
     $2\times10^{-8}$ & 7,231 & 1  \\
    \multicolumn{1}{c}{A large $k$ value} & \multirow{2}[0]{*}{7,232} & \multirow{2}[0]{*}{0}  \\
    \multicolumn{1}{c}{(no-equity-constraint case)} & & \\
    \hline
    \end{tabular}%
  \label{prevalence gap}%
\end{table}%

\end{appendices}

\renewcommand\refname{References} 
\bibliography{MultiSPReferences}

\end{document}